\documentclass[conference,letterpaper]{IEEEtran}
\IEEEoverridecommandlockouts
\usepackage[T1]{fontenc}
\usepackage{cite}
\usepackage{amsmath,amssymb,amsthm}
\usepackage{graphicx}
\usepackage{tikz}
\usetikzlibrary{arrows.meta,positioning,calc}
\usepackage{xcolor,colortbl}
\usepackage{booktabs,array}
\usepackage{algorithmic}
\usepackage{algorithm}
\usepackage{balance}
\usepackage{xurl}
\usepackage[hidelinks]{hyperref}
\graphicspath{{figs/}}
\theoremstyle{definition}
\newtheorem{definition}{Definition}
\theoremstyle{plain}
\newtheorem{theorem}{Theorem}
\newtheorem{lemma}{Lemma}

\newtheorem{remark}{Remark}

\newcommand{\MS}{\mathcal{M}_S} % small proxy model
\newcommand{\ML}{\mathcal{M}_L} % oracle model
\newcommand{\system}{CascadeFilter} % system name
\definecolor{AccentBlue}{RGB}{22,82,150}
\definecolor{RevisionBlue}{RGB}{0,70,180}
\definecolor{AccentLightBlue}{RGB}{232,241,252}
\definecolor{TableGray}{RGB}{246,247,249}

\newcommand{\oursrow}{\rowcolor{AccentLightBlue}}
\newcommand{\headrow}{\rowcolor{TableGray}}

\begin{document}
\title{CascadeFilter: Cost-Efficient Semantic Filtering with Joint Threshold Selection}
\author{
\IEEEauthorblockN{Yanji He$^{1}$, Tao Zhong$^{2}$, Hongyang Chen$^{3}$, Xiaobo Zhang$^{4}$, Mingyu Yang$^{1}$, and Wei Wang$^{1,5}$}
\IEEEauthorblockA{$^1$The Hong Kong University of Science and Technology (Guangzhou), Guangzhou, China;\\ 
$^2$University of New South Wales, Sydney, Australia; \\
$^3$Zhejiang Lab, Hangzhou, China; 
$^4$Southwest Jiaotong University, Chengdu, China; \\
$^5$The Hong Kong University of Science and Technology, Hong Kong SAR, China\\
    yhe720@connect.hkust-gz.edu.cn;
    tao.zhong@unswalumni.com;
    dr.h.chen@ieee.org; \\
    zhangxb@swjtu.edu.cn;
    myang250@connect.hkust-gz.edu.cn;
    weiwcs@ust.hk
    }
}
\maketitle

\begin{abstract}

Semantic filtering selects records that satisfy a natural-language predicate and is an important operation in data analytics based on large language models (LLMs). Executing an expensive model on every document incurs substantial inference cost. Existing approaches reduce this cost by accepting predictions from cheaper models and forwarding uncertain records to an oracle. Task Cascades further varies document portions and operations, and provides an accuracy-guaranteed extension. However, its threshold calibration follows a prescribed adjustment sequence after cascade assembly, rather than minimizing cost over alternative combinations of stage thresholds.

Motivated by this, we propose \system{}, a three-stage semantic filtering method that jointly searches the thresholds for a small language model operating on extracted evidence and full documents, with unresolved records forwarded to the oracle. For a fixed cascade, we formulate threshold selection as minimizing token-aware cost subject to an oracle-relative error constraint, and select the least-cost certified configuration in a pre-specified grid. A finite-population calibration procedure bounds oracle-relative error after selection, using fixed-sequence tests on threshold tracks to allocate the multiple-testing budget. We further combine lexical and dense retrieval, reranking, and neighborhood-supported confidence to improve evidence extraction and routing. These components enhance the cascade while leaving its certification interface unchanged. Across six datasets and 13 evaluation settings, the reported results have the highest benchmark accuracy and macro-averaged F1 score in 12 settings each. These comparisons describe the reported operating points; the accuracy guarantee concerns agreement with the oracle.
\end{abstract}

% =============================================================
% SECTION 1: INTRODUCTION
% =============================================================
{
\section{Introduction}
\label{sec:intro}

Semantic filtering identifies the records in a corpus that satisfy a natural-language predicate. Given a collection of documents and a query such as ``does this review mention a billing issue?'', the operation returns the documents for which the predicate is true. Large language models (LLMs) make such predicates available in data systems without requiring users to specify extraction rules or train a classifier for each query~\cite{patel2024lotus,liu2025palimpzest,shankar2024docetl}. Applications include selecting customer reviews by aspect, identifying dialogue outcomes, and checking whether textual evidence supports a claim. These applications require semantic decisions over large collections, making efficient execution of the filtering operation an important problem. Unlike a predicate over a structured attribute, a semantic predicate may depend on paraphrases, relationships between sentences, or the context in which a statement appears. A review can describe a billing problem without using that exact phrase, while a document containing those words need not report a problem. Retrieving a relevant passage helps identify evidence, but does not by itself replace the final semantic decision.

A direct execution strategy invokes a capable LLM on every document. Although simple, this strategy repeatedly pays for processing the entire input and generating a decision, even when a short passage is sufficient to answer the predicate. Its cost increases with both the number and length of documents. A cheaper small language model (SLM) can reduce inference cost, but accepting all its predictions can reduce accuracy. A model cascade therefore uses the cheaper model first and forwards uncertain records to a more expensive reference model, called the oracle. The objective is to reduce the total execution cost while meeting a user-specified accuracy target. Here accuracy is certified relative to the oracle; agreement with benchmark ground truth is evaluated separately because the oracle can itself make mistakes. Two decisions determine execution cost: how much text to process and which model to invoke. A short input can omit necessary evidence; a full document can supply that evidence without requiring a stronger model. Conversely, some records remain difficult even with full context. The execution policy should distinguish these cases and account for both input and output tokens.

Existing work addresses this cost--accuracy tradeoff through model routing, proxy prediction, and cascade execution. FrugalGPT and RouteLLM select among models with different costs and capabilities~\cite{chen2023frugalgpt,ong2024routellm}. LOTUS supports approximate semantic execution, while BARGAIN provides variants with statistical guarantees on accuracy, precision, or recall~\cite{patel2024lotus,10.1145/3769776}. ScaleDoc uses document representations and lightweight proxy models to reduce expensive predicate evaluations~\cite{zhang2026scaledoc}. These methods motivate using inexpensive computations to resolve easy records. Task Cascades extends this idea by varying the model, document portion, and operation within a cascade~\cite{shankar2026taskcascades}. Thus, exploiting shorter inputs and providing an accuracy guarantee are established directions; our focus is how to select coupled routing thresholds under an explicit cost objective.

\noindent\textbf{The State-of-the-art.}
Task Cascades~\cite{shankar2026taskcascades} generates candidate tasks using alternative models, document portions, and simplified or related operations. It first sets class-specific thresholds for individual tasks on development data, then greedily assembles a cascade from these tasks. Its accuracy-guaranteed extension subsequently adjusts the assembled cascade's thresholds on a validation split. In particular, it moves the thresholds together along a prescribed sequence of score-based offsets and tests the resulting cascade, stopping when certification fails~\cite{shankar2026taskcascades}. This procedure already accounts for cascade-level accuracy and should not be characterized as merely calibrating each stage independently. However, the guarantee adjustment is a post-processing step rather than a cost optimization over threshold combinations. Changing an early threshold changes both the number and composition of records reaching later stages. Consequently, different combinations can have different costs even when they meet the same accuracy target. This motivates jointly searching thresholds for a fixed cascade structure and comparing certified candidates by their total token cost. For example, relaxing the first threshold accepts additional records cheaply, but changes the error budget available downstream. Tightening the second threshold can compensate by forwarding more records to the oracle. Whether this exchange saves cost depends on which records move between stages and their processing costs. Evaluating each threshold separately cannot in general determine the best exchange. This comparison concerns threshold selection; our fixed structure does not subsume the broader task-generation capabilities of Task Cascades.

\noindent\textbf{Our Idea.}
We propose \system{}, which processes each record using an SLM on extracted evidence, then the SLM on the full document if needed, and finally the oracle for unresolved records. We formulate cost-minimizing threshold selection for this fixed cascade. The objective and feasible set are generally nonconvex, but error decreases with the second-stage threshold when the first is fixed. This structure motivates searching the two acceptance thresholds jointly over a grid fixed before calibration. For each pair, we account for errors across the complete cascade and certify oracle-relative accuracy using a sample drawn without replacement from the deployment population. We organize candidates into monotone tracks and apply fixed-sequence testing within each track, allocating the failure budget across tracks. We then select the least-cost certified candidate; this need not recover the true feasible optimum even within the grid. To improve the inputs and routing scores, we combine Best Matching 25 (BM25) lexical retrieval~\cite{robertson2009bm25} with dense retrieval using reciprocal rank fusion (RRF), rerank the resulting chunks, and incorporate embedding-neighborhood support into model confidence. These established retrieval components support the threshold-search design rather than constitute separate claims of new retrieval algorithms. The resulting method couples cost-aware selection with finite-population certification while retaining flexibility in evidence extraction and routing. The candidate grid and routing scores are frozen before calibration, preventing calibration labels from being reused to construct new policies. For each fixed first-stage threshold, second-stage candidates form a conservative-to-lenient sequence, allowing testing to stop at the first uncertified candidate. Searching across tracks retains alternative first-stage choices. If no candidate certifies, the method uses the oracle. The guarantee applies to the specified population and requires fresh calibration for deployment on a new population.

\noindent\textbf{Contributions.} We summarize our contributions as follows:
\begin{itemize}
\item \textbf{Joint Threshold Selection (Section~\ref{sec:system}).} We develop a three-stage semantic filtering method and jointly search its coupled thresholds, accounting for both context length and model cost. The search returns the least-cost certified configuration in a pre-specified grid.
\item \textbf{Finite-Population Certification (Section~\ref{sec:guarantee}).} We instantiate a without-replacement betting bound for cascade error and combine monotone tracks with fixed-sequence testing to control oracle-relative error after threshold selection. The betting method and fixed-sequence testing are existing tools; our contribution is their application to the coupled search.
\item \textbf{Evidence and Routing Enhancements (Section~\ref{sec:system}).} We integrate hybrid retrieval, reranking, batched evidence inference, and neighborhood-supported confidence into the cascade. These components improve the candidate policies evaluated by the certification procedure.
\item \textbf{Experimental Analysis (Section~\ref{sec:experiments}).} We evaluate six datasets with 13 settings and study cascade structure, routing components, and calibration behavior. The main table reports the highest reported benchmark accuracy and macro-averaged F1 score in 12 settings each, at the evaluated operating points.
\end{itemize}

\begin{figure}[!t]

\centering
\includegraphics[width=\linewidth]{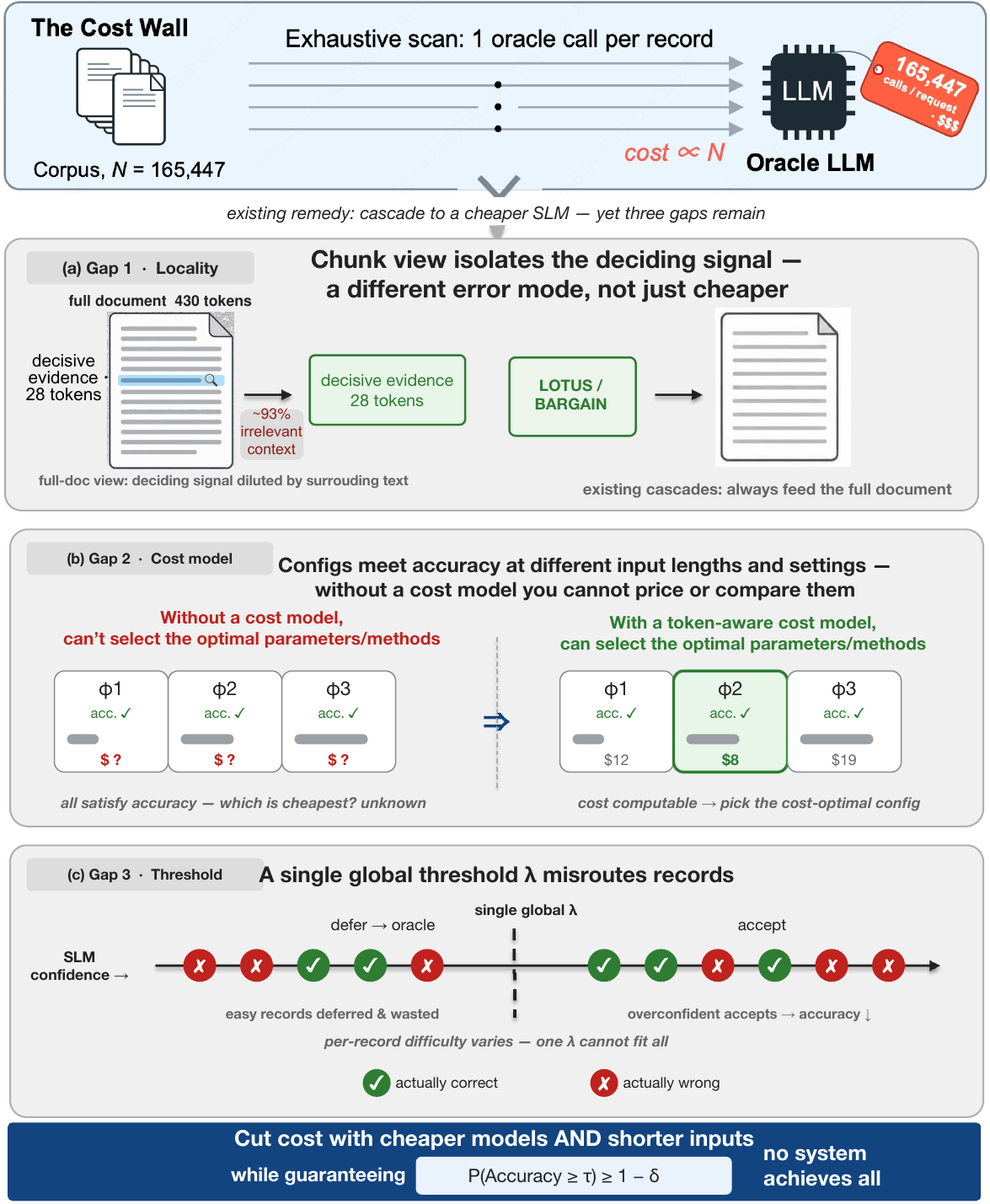}
\caption{Motivating examples: (a) compact evidence, (b) token-aware cost, and (c) routing by record difficulty. The labels illustrate these scenarios, not limitations of all prior cascades; numerical values are illustrative. Panel (c) motivates record-dependent scores, while the selected thresholds are shared across records.}
\label{fig:motivation}
\end{figure}

\noindent Due to space limitations, proofs and supplementary details are provided in Appendix~\ref{app:technical-report}, which serves as the technical report for this version. Code is available at \url{https://github.com/oceanusXXD/cascade}.

}

% =============================================================
% SECTION 2: PRELIMINARIES
% =============================================================
{
\section{Preliminaries}
\label{sec:problem}
 
\subsection{Task: Semantic Filtering}
\label{sec:prelim:task}
 
Semantic filtering selects the documents in a corpus that satisfy a
natural-language predicate. Given a query $q$ and $N$ documents
$D = \{d_1, \ldots, d_N\}$, the task returns
$D^* = \{d \in D : f^*(q, d) = 1\}$, where $f^*(q,d)\in\{0,1\}$ indicates
whether $d$ satisfies $q$.
Predicates such as \textit{``does this article support the given claim?''}
are semantic judgments that neither keyword matching nor embedding similarity
can reliably make~\cite{patel2024lotus}.

The ground truth $f^*$ is the evaluation target but is unavailable at corpus
scale. We therefore drive the cascade with a frontier LLM $\ML$ (e.g.\ GPT-5.5)
as a reference model, the \emph{oracle}: its disagreement with $f^*$ is accounted for separately below, while all
headline accuracies are against $f^*$. Served through an application programming interface (API) returning a
decision rather than calibrated probabilities, $\ML$ is used only as a binary
label $\ML(q,d)\in\{0,1\}$. We further assume an open-source small language
model (SLM) $\MS$ exposing token-level log probabilities for confidence-based
routing.

\subsection{Problem Definition}
\label{sec:prelim:problem}
We fix the cascade structure, models, evidence extractor, and routing-score functions. Stage~1 applies the SLM to extracted evidence, Stage~2 applies it to the full document, and Stage~3 invokes the oracle. The variables to optimize are the two acceptance thresholds $\phi=(\lambda_1,\lambda_2)$. For record $i\in\mathcal I=\{1,\ldots,N\}$, let $\tilde z_i^{(s)}$ and $\hat y_i^{(s)}$ denote its fixed score and prediction at stage $s\in\{1,2\}$. Define
\[
 D_i^{(s)}(\lambda_s)=\mathbf1[\tilde z_i^{(s)}<\lambda_s].
\]
A record stops at the first stage that accepts it. Stage~2 predictions need only be computed for records that can reach that stage under an admissible threshold.

\noindent\textbf{Execution cost.}
Let $c_{si}\geq0$ be the fixed processing cost of record $i$ at stage $s\in\{1,2,L\}$, with fixed batching at Stage~1. A token-based cost is $c_{si}=p_s^{\rm in}t_{si}^{\rm in}+p_s^{\rm out}\bar t_{si}^{\rm out}$, where $p_s$ denotes token prices, $t_{si}^{\rm in}$ the input length, and $\bar t_{si}^{\rm out}$ an output-length estimate fixed without calibration labels. Thus this objective is a modeled execution cost, not necessarily an observed invoice. With $C_1=\sum_i c_{1i}$,
\begin{equation}
\begin{aligned}
 C(\phi)=\mathrm{Cost}(\phi)
 &=C_1+\sum_i c_{2i}D_i^{(1)}\\
 &\quad+\sum_i c_{Li}D_i^{(1)}D_i^{(2)}.
\end{aligned}
\label{eq:cost}
\end{equation}
The first threshold controls which records incur second-stage cost; both thresholds jointly determine oracle cost. Per-record costs retain variation in document lengths rather than assuming that routed records have the population-average length.

\noindent\textbf{Oracle-relative error.}
Let $y_i=\ML(q,d_i)$, $a_i=\mathbf1[\hat y_i^{(1)}\neq y_i]$, and $b_i=\mathbf1[\hat y_i^{(2)}\neq y_i]$. Since oracle fallback has zero error relative to $y_i$, the population error is
\begin{equation}
 E(\phi)=\frac1N\sum_i\bigl[(1-D_i^{(1)})a_i
 +D_i^{(1)}(1-D_i^{(2)})b_i\bigr].
 \label{eq:population_error}
\end{equation}
Unlike routing and modeled cost, this quantity requires oracle labels.

\begin{definition}[Cost-Minimizing Threshold Selection]
\label{def:problem}
Given the fixed three-stage cascade, population $\mathcal I$, a candidate grid $\Phi=\Lambda_1\times\Lambda_2$, and target accuracy $\tau\in(0,1)$, find
\begin{equation}
 \phi^\star\in\arg\min_{\phi\in\Phi}C(\phi)
 \quad\text{subject to}\quad E(\phi)\leq\varepsilon=1-\tau.
 \label{eq:problem}
\end{equation}
If the grid has no feasible pair, use the all-oracle fallback, whose oracle-relative error is zero. Models, stage order, retrieval, and score functions are not optimization variables in this definition.
\end{definition}

\subsection{Nonconvexity and Threshold Structure}
\label{sec:prelim:structure}
To describe the geometry, first allow thresholds to vary continuously over a rectangle; the actual search restricts them to $\Phi$. Convexity refers to these continuous functions and their feasible set, rather than to the finite grid itself.

\begin{theorem}[Nonconvexity and partial monotonicity]
\label{thm:threshold_structure}
For fixed scores, predictions, and nonnegative per-record costs, $C$ and $E$ are constant on regions inducing the same routing decisions. In general, $C$ is not convex and $\{\phi:E(\phi)\leq\varepsilon\}$ is not convex. Nevertheless, $C$ is nondecreasing in either threshold, and, for fixed $\lambda_1$, $E$ is nonincreasing in $\lambda_2$. Error need not be monotone in $\lambda_1$.
\end{theorem}
\noindent The proof is given in Appendix~\ref{app:proof:threshold_structure}.

\noindent\textbf{Implications for search.}
For each fixed $\lambda_1$, the feasible Stage~2 thresholds form an upper interval, possibly empty, or a suffix of the ascending grid. The smallest feasible grid threshold minimizes cost on that track, possibly with ties. Hence, if population errors were available, enumerating first-stage thresholds and locating each track's feasibility boundary would suffice. With $N$ records there are at most $(N+1)^2$ distinct threshold-induced acceptance configurations, and direct evaluation on a grid of $L_1L_2$ candidates takes $O(NL_1L_2)$ time. Nonconvexity therefore does not imply hardness of unrestricted cascade assembly for this fixed two-threshold problem. The central difficulty is finding a low-cost feasible pair without labeling the entire population with the oracle.

\subsection{Selection with Limited Oracle Labels}
\label{sec:prelim:selection}
We draw a simple random sample without replacement (SRSWOR), $S\subseteq\mathcal I$ with $|S|=n$, and construct a certified set $\mathcal C(S)\subseteq\Phi$. The population, grid, predictions, scores, and modeled cost are fixed independently of the calibration draw. The sample retains its uniform random draw order. The algorithm returns $\hat\phi=A(S)\in\arg\min_{\phi\in\mathcal C(S)}C(\phi)$, with all-oracle fallback when this set is empty. Its statistical requirement is
\begin{equation}
 \Pr_S\{E(A(S))\leq\varepsilon\}\geq1-\delta.
 \label{eq:selection_guarantee}
\end{equation}
This requirement concerns the sample-dependent selection algorithm. For a fixed threshold pair, population error is deterministic. Certification may exclude truly feasible pairs, so the minimum over $\mathcal C(S)$ need not equal the optimum in Definition~\ref{def:problem}. We establish selection validity and minimum cost within the certified set, not an approximation ratio or recovery of $\phi^\star$. The monotonicity of true error motivates tracks; it does not itself imply monotonicity of the sample-based certification statistic or justify binary search with repeated tests.

The guarantee is relative to the oracle. For benchmark truth $f^*$,
\begin{equation}
 \mathrm{Acc}_{f^*}(\phi)\geq\mathrm{Acc}_{\ML}(\phi)-\Delta,
 \label{eq:gt_bridge}
\end{equation}
where $\Delta=N^{-1}\sum_i\mathbf1[y_i\neq f^*(q,d_i)]$. Thus a ground-truth guarantee requires an upper bound on oracle disagreement; a measured signed accuracy gap alone does not provide one.

Selected-plan cost excludes search overhead. Appendix~\ref{app:cost} details same-batch accounting and prediction reuse.

Table~\ref{tab:notation} in the appendix summarizes the notation.

\subsection{Calibration Principle}
\label{sec:prelim:wsr}
The deployment corpus is a fixed finite population; only the calibration sample is random. We use the Waudby-Smith and Ramdas (WS\&R) betting framework~\cite{waudbysmith2024} to test cascade error without labeling the entire corpus. Because labels are sampled without replacement, the test accounts for the remaining population after each draw. Section~\ref{sec:guarantee} explains how these tests remain valid when selecting among threshold pairs; Appendix~\ref{app:betting} gives the construction.

}

% =============================================================
% SECTION 3: SYSTEM DESIGN
% =============================================================
{
\section{\system{}: System Design}
\label{sec:system}
\subsection{Overview}
\label{sec:system:overview}
\system{} separates preparation, threshold selection, and execution (Algorithm~\ref{alg:cascade}). Preparation computes and caches the two small-model predictions and routing scores. Selection uses one oracle-labeled sample to find a low-cost certified threshold pair. Execution returns the first accepted small-model answer, or the oracle answer if both stages reject. The models, retrieval rules, and scores are fixed before calibration; only the two thresholds are selected.

\begin{figure*}[!t]

 \centering
 \includegraphics[width=\linewidth]{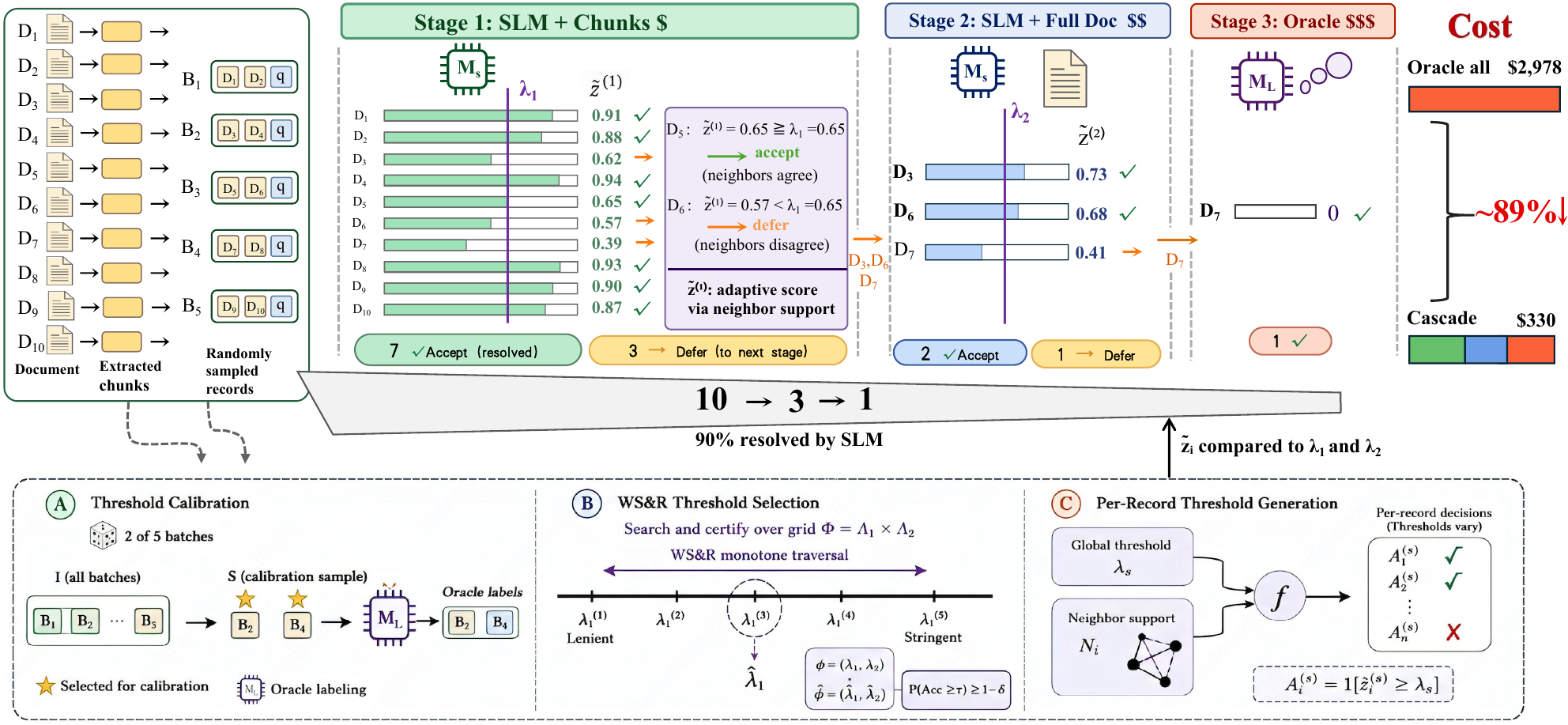}
 \caption{Three-stage overview: evidence and full-document SLM stages precede oracle fallback. Dollar values and routing counts are illustrative, not measured. Panel C adapts scores per record; the selected thresholds are shared. Calibration selects the least-cost certified pair, with oracle-relative validity under Theorem~\ref{thm:guarantee}; it need not certify every grid point.}

 \label{fig:architecture}
\end{figure*}
\begin{algorithm}[t]
\caption{\system{}: Prepare, Select, and Execute}
\label{alg:cascade}
\begin{algorithmic}[1]

\REQUIRE Documents, predicate, fixed models and score rules; $\tau,\delta,n$
\ENSURE One prediction per record
\STATE Extract evidence; cache Stage~1 predictions and scores.
\STATE Form $\Lambda_1$; identify records any candidate can defer.
\STATE Cache Stage~2 predictions and scores for these records.
\STATE Form $\Lambda_2$; freeze the grid $\Phi$ and candidate costs.
\STATE Sample $n$ records without replacement in uniform random order; obtain oracle labels.
\STATE Select $\hat\phi$ with Algorithm~\ref{alg:certify}.
\IF{no pair certifies}
\STATE Return oracle answers for all records; stop.
\ENDIF
\FOR{each record $i$}
\IF{Stage~1 score $\geq\hat\lambda_1$}
\STATE Return its cached Stage~1 prediction.
\ELSIF{Stage~2 score $\geq\hat\lambda_2$}
\STATE Return its cached Stage~2 prediction.
\ELSE
\STATE Return the oracle label, reusing it if already cached.
\ENDIF
\ENDFOR
\end{algorithmic}
\end{algorithm}

\subsection{Three Processing Stages}
\label{sec:system:stage1}
\noindent\textbf{Stage~1: short evidence and the small model.}
We split a document into overlapping sentence windows, combine BM25 and dense retrieval using RRF, and rerank the candidates to retain compact evidence. Several query--evidence pairs are processed in one small-model call. Each record receives its own prediction and score. A record is accepted when its score reaches $\lambda_1$; malformed outputs are deferred. Window sizes, batching, and score extraction are specified in Appendix~\ref{app:implementation}.

\noindent\textbf{Stage~2: full document and the small model.}
A record rejected by Stage~1 is reconsidered with the complete document. It is accepted if its new score reaches $\lambda_2$. During search, we precompute this result for every record that any candidate first-stage threshold could defer. All candidate pairs reuse these predictions; changing a threshold does not rerun the model.

\noindent\textbf{Stage~3: oracle fallback.}
A record rejected by both stages is answered by the oracle. Calibration labels are reused when available. This stage has zero error relative to the oracle by definition, but the oracle can disagree with ground truth. Modeled selected-plan cost and actual same-batch cost with prediction reuse are distinguished in Eq.~\eqref{eq:batch_cost}.

\subsection{Routing Scores}
\label{sec:system:neighbor_support}
The routing score combines the model's own confidence with agreement among nearby records in embedding space. Similar neighbors with matching predictions increase the score; disagreement lowers it. Sparse or distant neighborhoods receive less weight. This is a heuristic for ranking records by reliability, not an accuracy guarantee. The exact weighting and log-odds formula appear in Appendix~\ref{app:score}.

The statistical procedure only requires a fixed scalar score and acceptance rule $\tilde z_i^{(s)}\geq\lambda_s$. Neighbor predictions may be used to build the score, but calibration oracle labels may not. Keeping scores fixed preserves the nested acceptance sets needed by threshold search.

\subsection{Joint Threshold Selection}
\label{sec:system:joint_search}
We build a finite grid from score quantiles and evaluate complete threshold pairs. Changing the first threshold changes which records reach the second stage, so independent stage targets need not minimize total cost. Section~\ref{sec:guarantee} tests the combined cascade error, then chooses the cheapest certified pair. A pair may be truly feasible yet fail certification; the method therefore does not claim to recover the true feasible optimum.

\subsection{A Threshold-Selection Example}
\label{sec:system:example}
Consider a record whose fixed routing scores are $\tilde z^{(1)}=0.85$ and $\tilde z^{(2)}=0.90$. For illustration, suppose its stage costs are $c_1=1$, $c_2=3$, and $c_L=20$ in arbitrary units. The same cached predictions support three different execution paths:
\begin{center}
\small
\begin{tabular}{@{}lll@{}}
\hline
Threshold pair & Accepted at & Plan cost \\
\hline
$(0.80,0.85)$ & Stage~1 & $1$ \\
$(0.90,0.85)$ & Stage~2 & $1+3=4$ \\
$(0.90,0.95)$ & Oracle & $1+3+20=24$ \\
\hline
\end{tabular}
\end{center}
These costs are illustrative. Lowering the first threshold can avoid both later stages; changing the second affects only records that reach Stage~2. Candidate cost sums these path costs over the population.

Additional interpretation of this example appears in Appendix~\ref{app:threshold-example}.

}

% =============================================================
% SECTION 4: CERTIFIED THRESHOLD SEARCH
% =============================================================
{
\section{Certified Threshold Search}
\label{sec:guarantee}

We certify threshold pairs from one ordered calibration sample and select the cheapest certified pair. Certification takes $O(|\Phi|n)$ arithmetic operations on cached predictions; Theorem~\ref{thm:complexity} includes candidate-cost computation. The guarantee applies after selection and concerns oracle-relative error $\varepsilon=1-\tau$. All-oracle fallback is feasible but may be expensive.

For $\phi=(\lambda_1,\lambda_2)$ write
$A_i^{(s)}=\mathbf{1}[\tilde{z}_i^{(s)}\geq\lambda_s]$ and
$D_i^{(s)}=1-A_i^{(s)}$ for the stage-$s$ accept and defer indicators.
A record errs only if an SLM stage accepts it with a label differing from the oracle label
$y_i=\ML(q_i,d_i)$, Stage~3 being zero-loss by definition; the three
paths are mutually exclusive, so
\begin{equation}
 \begin{aligned}
 e_i(\phi)&=A_i^{(1)}\mathbf{1}[\hat{y}_i^{(1)}\!\neq\!y_i]
 \\ &\quad+\mathbf{1}[i\!\in\!\mathcal{I}_2^{\max}]\,D_i^{(1)}A_i^{(2)}\mathbf{1}[\hat{y}_i^{(2)}\!\neq\!y_i]
 \;\in\{0,1\}.
 \end{aligned}
 \label{eq:error_indicator}
\end{equation}
Writing $E(\phi)=\frac1N\sum_i e_i(\phi)$ for the population error rate, we have
$E(\phi)\leq\varepsilon\iff\mathrm{Acc}_{\ML}(\phi)\geq\tau$, so bounding $E(\phi)$ certifies
accuracy. The two routing rates that price a candidate, $r_{\mathrm{reach2}}(\phi)=\frac1N\sum_{i\in\mathcal{I}_2^{\max}}D_i^{(1)}$ and $r_3(\phi)=\frac1N\sum_{i\in\mathcal{I}_2^{\max}}D_i^{(1)}D_i^{(2)}$, are computable exactly on the full population
without any oracle query.

\subsection{Bounding a Single Candidate}
\label{sec:guarantee:wsr}
For a fixed threshold pair, compute its disagreement with the oracle on the calibration sample. We use a without-replacement betting test~\cite{waudbysmith2024}, denoted $M_n^+(\varepsilon;\phi)$, to test whether population error exceeds $\varepsilon$. A candidate passes at level $\beta$ when $M_n^+(\varepsilon;\phi)\geq1/\beta$. Larger evidence is required for smaller $\beta$. Appendix~\ref{app:betting} specifies the statistic and its residual-population centering.

\begin{theorem}[Single-candidate coverage]
\label{thm:coverage}
 Condition on the fixed population and candidate family. For any candidate $\phi$ with $E(\phi)>\varepsilon$, under the uniformly ordered draw and finite predictable bets of Appendix~\ref{app:betting},
 $\mathbb{P}(M_n^+(\varepsilon)\geq1/\beta)\leq\beta$.
 Hence the probability of certifying a candidate whose accuracy falls below $\tau$ is at most $\beta$.
\end{theorem}
\noindent The proof is given in Appendix~\ref{app:proof:coverage}.

\subsection{Searching the Grid under Selection}
\label{sec:guarantee:multi}
Because $\hat{\phi}$ is chosen by its own calibration outcome, per-candidate
bounds do not suffice: the search must control the risk of \emph{selecting} an
infeasible configuration. Rather than pay a flat multiplicity price we
exploit the structure of $\Phi$. The error rate is monotone along $\lambda_2$ but
\emph{not} along $\lambda_1$, since relaxing $\lambda_1$ moves records to the chunk view, which may raise or lower their error depending
on the predicate's locality. We therefore organize $\Phi$ into $L_1$ \emph{tracks},
one per $\lambda_1$, ordering $\lambda_{2,1}>\cdots>\lambda_{2,L_2}$ within each.

\begin{lemma}[Monotonicity along a track]
\label{lem:monotone}
 Fix $\lambda_1$ and let $E_m=E(\lambda_1,\lambda_{2,m})$. Then $E_1\leq\cdots\leq E_{L_2}$.
\end{lemma}
\noindent The proof is given in Appendix~\ref{app:proof:monotone}.

\noindent A pre-specified order and stopping at the first failed test give a valid \emph{fixed-sequence} scan: test
from the most stringent $\lambda_{2,1}$ toward the most lenient, certify while
the test passes, and stop at the first failure, spending the \emph{same} level
$\eta_f$ at every step. Across tracks we split $\delta$ by Bonferroni, $\eta_f=\delta/L_1$.
The certified set $\mathcal{C}$ is the union of the $L_1$ certified prefixes. Monotonicity gives this ordering its cost--error interpretation; the first-error proof itself only requires a fixed order and stopping rule.

\begin{lemma}[Certified-set validity]
\label{lem:certified_set}
 For a fixed family of $L_1$ nonempty tracks, fixed within-track orders, valid marginal tests, and stopping at the first failed test, with $\eta_f=\delta/L_1$, $\;\mathbb{P}(\exists\phi\in\mathcal{C}:E(\phi)>\varepsilon)\leq\delta$.
\end{lemma}
\noindent The proof is given in Appendix~\ref{app:proof:certified_set}.

\noindent The error-budget saving is that one false-certification event, at the first infeasible index $m_k^\star$, covers an
entire track: the search spends $\delta/L_1$ per track instead of $\delta/(L_1L_2)$
per candidate, so refining $\Lambda_2$ costs no confidence budget at all ---
the reason the joint search can afford a fine grid. Flat grid-wide
Bonferroni remains valid with smaller per-test levels. Fixed-sequence validity does not require monotone errors, but nonmonotone tracks can lose feasible candidates after an early failure. Score parameters $\psi_s$ extend the grid to $\Lambda_1\times\Lambda_2\times\Psi$ with $L_1\to L_1|\Psi|$ throughout; as the budget scales as $1/|\Psi|$ we keep $|\Psi|\leq3$.

\begin{algorithm}[t]
\caption{Certified Joint Threshold Search}
\label{alg:certify}
\begin{algorithmic}[1]

\REQUIRE Grid $\Phi$, each track ordered $\lambda_{2,1}>\cdots>\lambda_{2,L_2}$; sample
 $S\sim\mathrm{SRSWOR}(\mathcal{I},n)$ with oracle labels $\{y_i\}_{i\in S}$;
 $\varepsilon=1-\tau$; failure budget $\delta$; label-free $\mathrm{Cost}(\cdot)$
\ENSURE Certified set $\mathcal{C}$, deployed $\hat{\phi}$
\STATE $\mathcal{C}\leftarrow\varnothing$;\quad $\eta_f\leftarrow\delta/L_1$
 \hfill $\triangleright$ Bonferroni across $L_1$ tracks
\FOR{each track $k=1,\ldots,L_1$ (one per $\lambda_{1,k}$)}
 \FOR{$m=1,\ldots,L_2$ (stringent $\to$ lenient)}
 \STATE $W\leftarrow M_n^+(\varepsilon;\lambda_{1,k},\lambda_{2,m})$
 \hfill $\triangleright$ $O(n)$ sweep, Thm.~\ref{thm:coverage}
 \IF{$W<1/\eta_f$}
 \STATE \textbf{break} \hfill $\triangleright$ fixed sequence, Lem.~\ref{lem:monotone}
 \ENDIF
 \STATE $\mathcal{C}\leftarrow\mathcal{C}\cup\{(\lambda_{1,k},\lambda_{2,m})\}$
 \ENDFOR
\ENDFOR
\STATE $\hat{\phi}\leftarrow\arg\min_{\phi\in\mathcal{C}}\mathrm{Cost}(\phi)$, or full Stage-3 fallback if $\mathcal{C}=\varnothing$
\RETURN $\mathcal{C},\ \hat{\phi}$
 \hfill $\triangleright$ $\mathbb{P}(\mathrm{Acc}_{\ML}(\hat{\phi})\geq\tau)\geq1-\delta$
\end{algorithmic}
\end{algorithm}

\subsection{End-to-End Guarantee}
\begin{theorem}[Composable accuracy guarantee]
\label{thm:guarantee}
 Assume (1)~the population, cached predictions, score family, grids, test rules, track orders, and $\delta$-allocation are fixed independently of the calibration draw; (2)~$1\leq n\leq N$ is fixed and $S$ retains its uniform without-replacement draw order; (3)~Stage~2 covers $\mathcal I_2^{\max}$ and the tests and scan satisfy Theorem~\ref{thm:coverage} and Lemma~\ref{lem:certified_set}; and (4)~the selected pair minimizes cost in $\mathcal C$, with all-oracle fallback if $\mathcal C=\varnothing$. Oracle fallback must return the same fixed labels $y_i$ that define the target. Then
 $\mathbb{P}(\mathrm{Acc}_{\ML}(\hat{\phi})\geq\tau)\geq1-\delta$, and
 when $\mathcal{C}\neq\varnothing$, $\hat{\phi}$ is the cheapest certified configuration.
\end{theorem}
\noindent The proof is given in Appendix~\ref{app:proof:guarantee}.

The guarantee applies after threshold selection and is relative to the oracle. The score determines which candidates can save cost; certification determines whether there is sufficient evidence to accept them. Extensions to other loss functions are discussed in Appendix~\ref{rem:bounded}.

\noindent\textbf{Cost-optimal selection.}
\label{sec:guarantee:cost}
Equation~\eqref{eq:cost} is a weighted sum of routing indicators, computable without oracle labels using fixed per-record cost estimates. After certification, scanning $\mathcal C$ returns its least-cost member. This does not establish minimum cost over all feasible members of $\Phi$: finite-sample conservatism and track stopping may omit feasible candidates. Costs need not be proportional to oracle-call counts when record lengths differ. Same-batch execution with reused predictions requires the separate accounting in Eq.~\eqref{eq:batch_cost}.

\begin{theorem}[Computational complexity]
\label{thm:complexity}
Let $N$ be the population size, $n\leq N$ the calibration size, and $G=L_1L_2\geq1$ the number of threshold pairs for one fixed score family. Suppose cached per-record routing and cost contributions can be evaluated in constant time and each betting update takes constant time. Using sorting to construct the quantile grids and a direct population scan to price each candidate, Algorithm~\ref{alg:cascade} requires
\[
 O\bigl(N\log N+G(N+n)\bigr)
\]
arithmetic operations and $O(N+G)$ auxiliary scalar storage. Its certification phase alone costs $O(Gn)$ time. These bounds exclude retrieval, score construction, model inference, and storage of documents, embeddings, and model parameters. If $R$ is the deployed oracle residual, caching requires $|S\cup R|\leq\min\{N,n+|R|\}$ oracle evaluations in total.
\end{theorem}
\noindent The proof is given in Appendix~\ref{app:proof:complexity}.

The total runtime additionally includes preprocessing and model execution, whose cost depends on document lengths, retrieval backbones, and batching. The algorithm evaluates Stage~1 on $N$ records and Stage~2 on at most $N$ records; these predictions are reused across all $G$ candidates. Thus the grid search multiplies scalar computation, rather than requiring a new model-inference pass per candidate. The theorem is an upper bound for the stated direct-scan algorithm, not an NP-hardness claim for the restricted threshold problem.

\noindent\textbf{Scope and assumptions.}
\label{sec:guarantee:scope}
The single-point certificate is conservative, and the claim holds within the family and procedure frozen before calibration (Theorem~\ref{thm:guarantee}, assumption~1): \system{} returns the least-cost certified candidate, not a global optimum. Exact $r_3,r_{\mathrm{reach2}}$ need one small-model Stage-2 pass over $\mathcal I_2^{\max}$; its computation is incurred once, and its cached predictions must not be charged again as selected-plan invocations. This distinction is explicit in Eq.~\eqref{eq:batch_cost}. Every statement is finite-population over the specific $\mathcal I$, with no i.i.d.\ assumption; a stream is handled by recalibrating per window from a fresh draw.

The theorem concerns one calibration run. Simultaneous validity across multiple deployments requires an additional allocation of failure probabilities whose sum is bounded by the overall budget.

}

% =============================================================
% SECTION 5: EXPERIMENTS
% =============================================================
{
\section{Experiments}
\label{sec:experiments}
Our evaluation addresses five research questions (RQs). \textbf{RQ1} (accuracy and cost): how do the reported operating points compare across methods? \textbf{RQ2} (cascade structure): how does removing either small-model stage affect the measured results? \textbf{RQ3} (components): how do Neighbor Support and evidence retrieval affect those operating points? \textbf{RQ4} (calibration): what do repeated draws reveal about certificate behavior? \textbf{RQ5} (sensitivity): how do the target accuracy, failure budget, sample size, and model backbones affect the selected policy? The results below are reported measurements; unresolved run-provenance and accounting issues are stated in Sections~\ref{sec:exp:validity} and~\ref{sec:exp:threats}.

\subsection{Experimental Setup}
\label{sec:exp:setup}
\noindent\textbf{Datasets and predicates.}
We evaluate \system{} on six datasets spanning 13 evaluation settings and three document genres (Table~\ref{tab:datasets}), each classified along the two axes the design targets. \emph{Locality} (Loc.) asks whether a single retrievable span decides the predicate (H), whether the answer is local but needs corroboration (M), or whether evidence is spread across the document (L); this is the axis Stage~1 exploits (Figure~\ref{fig:motivation}a). \emph{Balance} (Bal.) marks a rare positive class (skew), the regime an accuracy-only target leaves unpoliced. Both labels are assigned a priori, from the predicate wording alone and before any run. Since most public suites ship only a handful of fixed queries, we build the Yelp benchmark ourselves: 50 hand-authored aspect queries over 2,000 cleaned Yelp texts form 100K $(q,d)$ records. The finite-population argument permits certification over this pooled population without a homogeneity assumption. Ground truth is the consensus of three independent labelers (Claude Sonnet~4.6, ChatGLM~5.2, DeepSeek-V4-Pro), keeping only records on which all three agree; on a random 1K sample these match human annotation at Cohen's $\kappa=0.94$ (98.5\% raw agreement). The cited workload uses model-generated reference labels for Codebase and RV-Q2/Q3; their provenance in our runs needs confirmation. The public code supplies a three-record example, not the full benchmark or experiment traces.

\begin{table}[t]

\caption{Evaluation workload (13 settings; Yelp aggregates 50 aspect templates).
Non-Yelp predicate definitions follow~\cite{hou2026beyondlinear}.
Loc.: qualitative a priori evidence locality (H/M/L), not a measured context requirement.
Bal.: class-balance annotation; skew: rare positive class;
--: not flagged as rare-positive.}
\label{tab:datasets}
\centering
\footnotesize
\setlength{\tabcolsep}{2pt}
\renewcommand{\arraystretch}{1.0}

\begin{tabular}{@{}
    p{0.14\columnwidth}
    p{0.61\columnwidth}
    c c @{}}
\hline
\textbf{ID} & \textbf{Predicate} &
\textbf{Loc.} & \textbf{Bal.} \\
\hline

\multicolumn{4}{@{}l}{\textbf{Yelp (ours)}} \\
Yelp &
50 aspect templates (e.g., complaints about wait time) &
M & -- \\[1pt]

\multicolumn{4}{@{}l}{\textbf{Codebase}~\cite{li2023bird}} \\
CB-Q1 &
Does the user's $\langle\mathrm{AboutMe}\rangle$ contain a social-media link? &
H & skew \\
CB-Q2 &
Does $\langle\mathrm{AboutMe}\rangle$ indicate experience in or interest in computer science? &
M & -- \\
CB-Q3 &
Does $\langle\mathrm{AboutMe}\rangle$ focus on factual identification details? &
L & -- \\[1pt]

\multicolumn{4}{@{}l}{\textbf{IMDB-Review}~\cite{maas2011learning}} \\
RV-Q1 &
Is the overall review sentiment positive? &
M & -- \\
RV-Q2 &
Does the review explicitly recommend that others watch the movie? &
H & -- \\
RV-Q3 &
Does the review focus on factual descriptions, such as technical details? &
L & skew \\[1pt]

\multicolumn{4}{@{}l}{\textbf{AirDialogue}~\cite{wei2018airdialogue}} \\
AD-Q1 &
Does the dialogue have outcome $\langle\mathrm{book}\rangle$? &
M & -- \\
AD-Q2 &
Does the dialogue have outcome $\langle\mathrm{cancel}\rangle$? &
M & skew \\
AD-Q3 &
Does the dialogue have outcome $\langle\mathrm{no\_flight}\rangle$? &
M & -- \\
AD-Q4 &
Does the dialogue have outcome $\langle\mathrm{no\_reservation}\rangle$? &
M & -- \\[1pt]

\multicolumn{4}{@{}l}{\textbf{Twitter Hate}~\cite{davidson2017automated}} \\
TC &
Does the tweet contain hate speech or offensive language? &
M & -- \\[1pt]

\multicolumn{4}{@{}l}{\textbf{FEVER}~\cite{thorne2018fever}} \\
FEVER &
Is the claim supported by the provided evidence? &
M & -- \\

\hline
\end{tabular}
\end{table}

\noindent\textbf{Baselines.}
We compare seven baselines. \textbf{Full SLM} applies the small model to every record. \textbf{LOTUS}~\cite{patel2024lotus} uses proxy-based semantic execution; \textbf{BARGAIN}~\cite{10.1145/3769776} selects a proxy--oracle routing threshold with quality guarantees. \textbf{UniVote/SimVote} denote the uniform and similarity-weighted voting variants of Clustering-Sampling-Voting (CSV)~\cite{hou2026beyondlinear}, which infer tuple labels from sampled cluster members and support semantic filtering. \textbf{ScaleDoc}~\cite{zhang2026scaledoc} trains a query-specific proxy on precomputed document representations. \textbf{Task Cascades +G} adds accuracy-guaranteed threshold adjustment to an assembled task cascade~\cite{shankar2026taskcascades}. BARGAIN supports accuracy, precision, and recall variants, and Task Cascades +G provides an oracle-relative accuracy guarantee. These targets differ from the benchmark-label metrics reported here. The reported comparisons use common small and oracle models at one operating point per method; baseline-specific parameters, prompts, label budgets, and source revisions still require verification from run manifests. The tables therefore describe these operating points and do not establish performance at matched accuracy or dominance over each baseline's attainable tradeoff curve.

\noindent\textbf{Models and calibration.}
$\MS$ uses Qwen3-4B-Instruct-2507~\cite{qwen2025qwen3_4b_instruct_2507} ($c_S^{\mathrm{in}}{=}c_S^{\mathrm{out}}{=}\$0.2$ per million tokens) for inference, Qwen3-Embedding-4B~\cite{qwen3embedding_2025} for retrieval, and Qwen3-Reranker-0.6B~\cite{qwen3embedding_2025} as the cross-encoder, reported as running on a single A100-80GB graphics processing unit (GPU). Stage~3 dispatches to the frontier API model $\ML$, GPT-5.5~\cite{openai2026gpt55api} ($c_L^{\mathrm{in}}{=}\$5,c_L^{\mathrm{out}}{=}\$30$ per million tokens), $25$--$150\times$ costlier per token than $\MS$, with thinking mode disabled to minimize its cost. The self-hosted SLM is assigned a nominal rate of $\$0.2$ per million input or output tokens. This normalizes the token-cost comparison; it is not a measured GPU operating cost or a verified provider tariff. The stated oracle rates match its documented standard text-token rates. Temperature is set to zero; the theorem additionally requires fixed potential outputs independent of calibration sampling (Appendix~\ref{app:betting}). Two structural hyperparameters are fixed across all 13 settings and never tuned per predicate: the Stage-1 batch token budget $B_{\max}{=}8,192$ (Section~\ref{sec:system:stage1:routing}), packing 20--40 records per call at the observed evidence lengths, and the neighborhood size $K{=}20$ (Section~\ref{sec:system:neighbor_support}). Accuracy and Macro-F1 are against benchmark reference labels; the oracle is invoked only for Stage~3 fallback and for labeling $S$. Unless noted, all main-table results use $\tau{=}0.9$, $\delta{=}0.05$, and $n=\max\{\mathrm{ROUND\_HALF\_UP}(0.01N),n_0(N,\tau,\delta)\}$, with the necessary zero-error feasibility floor $n_0$ discussed in Section~\ref{sec:exp:validity}.

\noindent\textbf{Certification grid.}
The grids are empirical quantiles of the frozen routing scores, formed before $S$: up to 32 points for $\Lambda_1$ and 64 for $\Lambda_2$ (duplicates removed), so $|\Phi|\leq2048$, with $|\Psi|{=}1$ and $\eta_f=\delta/L_1$ per track. Validity holds at any resolution; resolution and $\delta$-allocation shift only how many lenient configurations certify, hence the realized $r_3$. Label budgets run from 119 to 3,618, so the smallest corpora bind on power (Section~\ref{sec:exp:validity}). For certificate validation, we repeat calibration and deployment over 20 independent draws of $S$ per setting, using the default calibration-size rule.

\noindent\textbf{Cost accounting.}
The reported cost model follows Eq.~\eqref{eq:cost}, using input-token and invocation counts and mean output lengths. The cost in Figure~\ref{fig:cost} is described as including all three stages, calibration labeling, the label-free Stage-2 search pass over $\mathcal I_2^{\max}$ (Section~\ref{sec:guarantee:scope}), and embedding, BM25, reranking, and indexing compute. These reported totals require reconciliation with Eq.~\eqref{eq:batch_cost} to verify that cached Stage-2 predictions and reused calibration labels are not charged twice; they should not yet be interpreted as audited end-to-end bills. The reported equal-weight mean cost across the 13 settings is $\$2.968\times10^{-4}$ per predicate--record. The token panel counts only oracle tokens.

\subsection{Main Results (RQ1)}
\label{sec:exp:main}
Table~\ref{tab:main_acc} reports Accuracy and Macro-F1 across all 13 evaluation settings. \system{} has the highest reported Accuracy and Macro-F1 in 12 of 13 settings, averaging 0.938/0.887; the only losses are on AD-Q4 (BARGAIN 0.902/0.866 vs.\ our 0.901/0.846), explained below. CascadeFilter's benchmark Accuracy standard deviations are 0.013 on AD-Q2 and 0.006 on AD-Q4 (Table~\ref{tab:routing}). The point differences on AD-Q2 (0.907 vs.\ 0.903 against Task Cascades) and AD-Q4 (0.901 vs.\ 0.902 against BARGAIN) are small. Per-draw spreads for those baselines are not reported, so these comparisons alone do not establish statistical significance. The comparisons below describe the table without attributing differences to a single mechanism.

\begin{table*}[!t]

\centering
\caption{Accuracy and Macro-F1 across 13 evaluation settings. Avg denotes the equal-weight arithmetic mean across the 13 settings. The best value in each column is highlighted.}
\label{tab:main_acc}
\scriptsize
\setlength{\tabcolsep}{2.4pt}
\renewcommand{\arraystretch}{0.84}

\resizebox{\textwidth}{!}{%
\begin{tabular}{@{}lcccccccccccccc@{}}
\toprule
\multicolumn{15}{@{}l}{\textbf{(a) Accuracy}} \\
\midrule
\textbf{Method}
& \textbf{Yelp}
& \textbf{CB-Q1}
& \textbf{CB-Q2}
& \textbf{CB-Q3}
& \textbf{FEVER}
& \textbf{TC}
& \textbf{RV-Q1}
& \textbf{RV-Q2}
& \textbf{RV-Q3}
& \textbf{AD-Q1}
& \textbf{AD-Q2}
& \textbf{AD-Q3}
& \textbf{AD-Q4}
& \textbf{Avg} \\
\midrule

Full SLM
& 0.743 & 0.766 & 0.679 & 0.614 & 0.694 & 0.625 & 0.826 & 0.720 & 0.736 & 0.749 & 0.731 & 0.704 & 0.720 & 0.716 \\

LOTUS
& 0.814 & 0.930 & 0.771 & 0.717 & 0.804 & 0.728 & 0.905 & 0.799 & 0.846 & 0.820 & 0.833 & 0.846 & 0.800 & 0.816 \\

BARGAIN
& 0.895 & 0.928 & 0.878 & 0.875 & 0.892 & 0.882 & 0.913 & 0.885 & 0.901 & 0.875 & 0.896 & 0.898 & \textbf{0.902} & 0.894 \\

ScaleDoc
& 0.840 & 0.881 & 0.848 & 0.804 & 0.838 & 0.849 & 0.841 & 0.768 & 0.870 & 0.800 & 0.820 & 0.752 & 0.823 & 0.826 \\

UniVote
& 0.738 & 0.856 & 0.700 & 0.633 & 0.668 & 0.668 & 0.809 & 0.689 & 0.851 & 0.821 & 0.726 & 0.712 & 0.729 & 0.738 \\

SimVote
& 0.757 & 0.873 & 0.704 & 0.661 & 0.666 & 0.699 & 0.791 & 0.664 & 0.814 & 0.824 & 0.733 & 0.719 & 0.739 & 0.742 \\

Task Cascades
& 0.898 & 0.912 & 0.908 & 0.877 & 0.899 & 0.887 & 0.911 & 0.915 & 0.891 & 0.899 & 0.903 & 0.896 & 0.899 & 0.900 \\

\oursrow\system{}
& \textbf{0.943} & \textbf{0.975} & \textbf{0.956} & \textbf{0.951} & \textbf{0.943} & \textbf{0.931} & \textbf{0.955} & \textbf{0.937} & \textbf{0.947} & \textbf{0.931} & \textbf{0.907} & \textbf{0.919} & 0.901 & \textbf{0.938} \\

\midrule
\addlinespace[2pt]
\multicolumn{15}{@{}l}{\textbf{(b) Macro-F1}} \\
\midrule
\textbf{Method}
& \textbf{Yelp}
& \textbf{CB-Q1}
& \textbf{CB-Q2}
& \textbf{CB-Q3}
& \textbf{FEVER}
& \textbf{TC}
& \textbf{RV-Q1}
& \textbf{RV-Q2}
& \textbf{RV-Q3}
& \textbf{AD-Q1}
& \textbf{AD-Q2}
& \textbf{AD-Q3}
& \textbf{AD-Q4}
& \textbf{Avg} \\
\midrule

Full SLM
& 0.629 & 0.612 & 0.647 & 0.580 & 0.687 & 0.537 & 0.823 & 0.678 & 0.519 & 0.736 & 0.468 & 0.665 & 0.614 & 0.630 \\

LOTUS
& 0.754 & 0.774 & 0.769 & 0.701 & 0.802 & 0.653 & 0.905 & 0.781 & 0.630 & 0.820 & 0.522 & 0.807 & 0.687 & 0.739 \\

BARGAIN
& 0.884 & 0.776 & 0.876 & 0.854 & 0.892 & 0.784 & 0.913 & 0.868 & 0.631 & 0.875 & 0.543 & 0.865 & \textbf{0.866} & 0.817 \\

ScaleDoc
& 0.824 & 0.672 & 0.847 & 0.736 & 0.838 & 0.732 & 0.840 & 0.710 & 0.592 & 0.800 & 0.512 & 0.707 & 0.779 & 0.738 \\

UniVote
& 0.623 & 0.534 & 0.698 & 0.613 & 0.656 & 0.553 & 0.808 & 0.674 & 0.487 & 0.819 & 0.467 & 0.637 & 0.616 & 0.630 \\

SimVote
& 0.649 & 0.494 & 0.700 & 0.641 & 0.653 & 0.598 & 0.790 & 0.648 & 0.506 & 0.823 & 0.471 & 0.642 & 0.626 & 0.634 \\

Task Cascades
& 0.889 & 0.741 & 0.906 & 0.846 & 0.899 & 0.789 & 0.910 & 0.903 & 0.704 & 0.898 & 0.579 & 0.862 & 0.865 & 0.830 \\

\oursrow\system{}
& \textbf{0.934} & \textbf{0.905} & \textbf{0.956} & \textbf{0.941} & \textbf{0.943} & \textbf{0.892} & \textbf{0.955} & \textbf{0.926} & \textbf{0.809} & \textbf{0.931} & \textbf{0.594} & \textbf{0.896} & 0.846 & \textbf{0.887} \\

\bottomrule
\end{tabular}%
}
\end{table*}

\noindent\textbf{Accuracy and oracle routing.}
On CB-Q3, Full SLM reaches 0.614 accuracy and the cascade reaches 0.951, a difference of 33.7 percentage points (pp), with 29.9\% of records assigned to Stage~3. The cascade also changes the input to Stage~1 and its routing scores, so this difference cannot be attributed to oracle calls alone. Stage~3 routing is highest on RV-Q3 (31.2\%) and lowest on CB-Q1 (2.5\%). The corresponding Full-SLM accuracies are 0.736 and 0.766; Full SLM is not already near ceiling on CB-Q1. The stage-level error breakdown needed to isolate these effects is not reported.

\noindent\textbf{Routing and annotated evidence locality.}
Table~\ref{tab:datasets} labels CB-Q1 and RV-Q2 as high-locality (H), CB-Q3 and RV-Q3 as low-locality (L), and the remaining settings as medium-locality (M). From Table~\ref{tab:routing}, their mean Stage-1 acceptance shares are 69.1\%, 37.7\%, and 59.4\%, respectively. These descriptive averages are consistent with greater early acceptance for more local predicates, but do not isolate a causal effect. CB-Q1 asks about social-media links, and RV-Q2 asks whether a review recommends the movie. RV-Q1 and AD-Q1 are medium-locality settings in the workload definition. The oracle-routing rates range from 2.5\% to 31.2\%, with an equal-weight mean of 14.6\%.

\noindent\textbf{Differences at the reported operating points.}
UniVote and SimVote reach 0.726/0.733 Accuracy on AD-Q2 and 0.729/0.739 on AD-Q4. BARGAIN reaches 0.898 on AD-Q3, compared with 0.919 for \system{}. These comparisons do not identify the source of the differences: the systems differ in label propagation, calibration, and routing, in addition to their scores. Across these 13 settings, \system{}'s lowest reported Accuracy is 0.901 on AD-Q4. This observation does not establish robustness on other predicates or prove that Neighbor Support causes the improvement; the component-removal study provides the more direct comparison.

\noindent\textbf{Accuracy and class imbalance.}
The certificate controls overall oracle-relative error. It does not guarantee Macro-F1, precision, recall, or minority-class performance; Appendix~\ref{rem:bounded} explains why these require additional analysis. The three settings marked skew in Table~\ref{tab:datasets} are CB-Q1 (0.975/0.905 Accuracy/Macro-F1), RV-Q3 (0.947/0.809), and AD-Q2 (0.907/0.594). Their metric gaps illustrate why overall Accuracy alone is insufficient to describe class-wise performance. AD-Q4 is not marked skew in that table. It is the one setting where BARGAIN has higher reported Accuracy and Macro-F1 than \system{} (0.902/0.866 versus 0.901/0.846). No significance claim follows from these point estimates.

\noindent\textbf{Reported cost and quality.}
Figure~\ref{fig:cost} reports nominal cost, oracle-call counts, and oracle-token counts; Figure~\ref{fig:tradeoff} groups the measured operating points by dataset. The oracle has a higher assigned per-token price, but its contribution also depends on the lengths of the routed records. Full SLM, UniVote, and SimVote have lower reported cost and lower mean benchmark Accuracy (0.716, 0.738, and 0.742). These benchmark-label comparisons cannot determine whether an oracle-relative constraint is met. \system{} has higher reported mean Accuracy than the other baselines, but the cost ratios still require a common accounting ledger for preprocessing, calibration, caching, and deployment. We therefore do not infer a matched-accuracy advantage, an audited end-to-end saving, or a Pareto-frontier dominance result from these plots. The modeled-cost selection theorem concerns candidates within our fixed cascade, not optimality over other systems.

\begin{figure*}[!t]
\centering
\includegraphics[width=\textwidth]{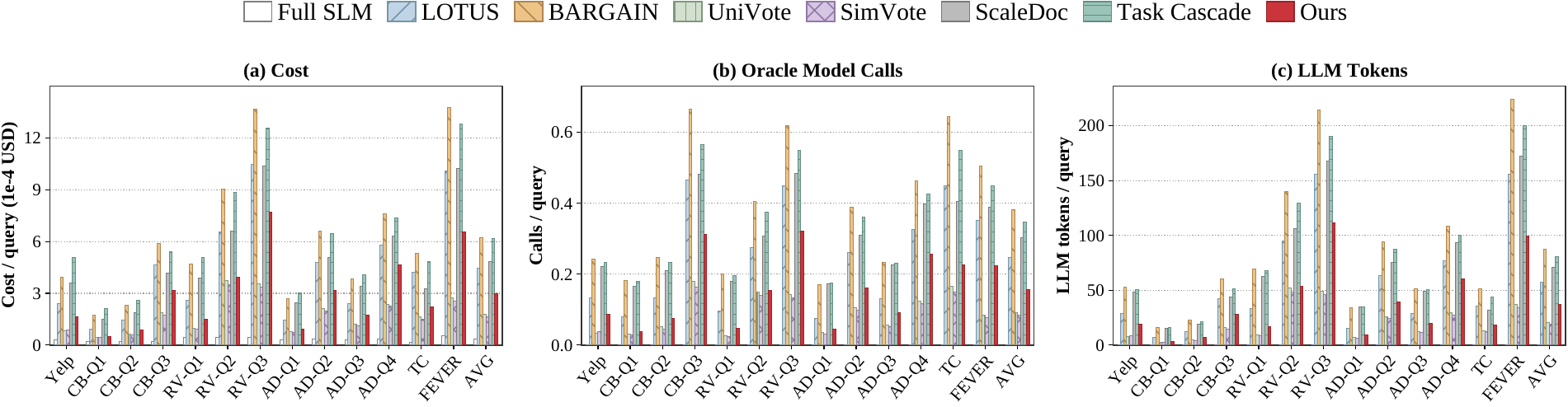}
\caption{Reported nominal cost, oracle calls, and oracle-token usage per predicate--record across 13 settings; cost accounting remains unaudited; \textsc{Avg} denotes the equal-weight arithmetic mean across the 13 settings.}

\label{fig:cost}
\end{figure*}

\begin{table}[!t]

\centering
\caption{Per-setting reference deployment and oracle-agreement profile of
CascadeFilter ($\tau=0.9$, $\delta=0.05$).
The routing split describes the reference operating point.
Acc is CascadeFilter's ground-truth accuracy;
$\mathrm{Acc}_{M_L}$ is its agreement with the oracle, not the
oracle's own ground-truth accuracy.
Both are shown as mean $\pm$ sample standard deviation over
20 realizations per setting.
$\widehat{\Delta}=\overline{\mathrm{Acc}}_{M_L}
-\overline{\mathrm{Acc}}$ is the signed difference between the means.
$V$ counts oracle-relative target violations among those realizations.
Mgn is unreported without corresponding wealth-test logs.
Statistics are computed before rounding.
The summary totals $N$ (predicate--record instances) and $V$;
its routing split is the unweighted setting mean, and its accuracy
statistics use the per-realization macro-averages over settings.}
\label{tab:routing}
\scriptsize
\setlength{\tabcolsep}{1.3pt}
\renewcommand{\arraystretch}{1.08}
\resizebox{\columnwidth}{!}{%
\begin{tabular}{@{}lrrrrrrrcc@{}}
\toprule
\headrow
\textbf{Setting} & \textbf{$N$} &
\multicolumn{3}{c}{\textbf{Routing split (\%)}} &
\textbf{Acc} & \textbf{$\mathrm{Acc}_{M_L}$} &
\textbf{$\widehat{\Delta}$} & \textbf{$V$} & \textbf{Mgn} \\
\cmidrule(lr){3-5}
& & \textbf{S1} & \textbf{S2} & \textbf{S3} & & & & & \\
\midrule
Yelp & 100K & 66.8 & 25.6 & 7.6 & $.943\pm.006$ & $.940\pm.006$ & -.002 & 0 & -- \\
CB-Q1 & 9.4K & 86.8 & 10.7 & 2.5 & $.975\pm.011$ & $.976\pm.011$ & .001 & 0 & -- \\
CB-Q2 & 9.4K & 61.2 & 32.7 & 6.1 & $.956\pm.004$ & $.963\pm.004$ & .007 & 0 & -- \\
CB-Q3 & 9.4K & 38.4 & 31.7 & 29.9 & $.951\pm.007$ & $.953\pm.007$ & .001 & 0 & -- \\
FEVER & 165.4K & 52.1 & 26.5 & 21.4 & $.943\pm.010$ & $.944\pm.010$ & .002 & 0 & -- \\
TC & 24.8K & 45.1 & 33.2 & 21.7 & $.931\pm.009$ & $.937\pm.010$ & .006 & 0 & -- \\
RV-Q1 & 50K & 82.6 & 13.6 & 3.8 & $.955\pm.004$ & $.959\pm.004$ & .003 & 0 & -- \\
RV-Q2 & 50K & 51.3 & 34.2 & 14.5 & $.937\pm.012$ & $.938\pm.012$ & .001 & 0 & -- \\
RV-Q3 & 50K & 36.9 & 31.9 & 31.2 & $.947\pm.011$ & $.948\pm.011$ & .001 & 0 & -- \\
AD-Q1 & 361.8K & 83.7 & 12.8 & 3.5 & $.931\pm.009$ & $.933\pm.009$ & .002 & 0 & -- \\
AD-Q2 & 361.8K & 51.6 & 33.2 & 15.2 & $.907\pm.013$ & $.907\pm.013$ & .000 & 6 & -- \\
AD-Q3 & 361.8K & 56.0 & 35.8 & 8.2 & $.919\pm.019$ & $.922\pm.020$ & .002 & 3 & -- \\
AD-Q4 & 361.8K & 35.9 & 39.5 & 24.6 & $.901\pm.006$ & $.910\pm.006$ & .009 & 0 & -- \\
\midrule
\textbf{Summary} & \textbf{1.9M} & \textbf{57.6} & \textbf{27.8} & \textbf{14.6} & $\mathbf{.938\pm.003}$ & $\mathbf{.941\pm.003}$ & \textbf{.003} & \textbf{9/260} & -- \\
\bottomrule
\end{tabular}%
}
\end{table}

\begin{figure}[!t]

\centering
\includegraphics[width=0.9\columnwidth]{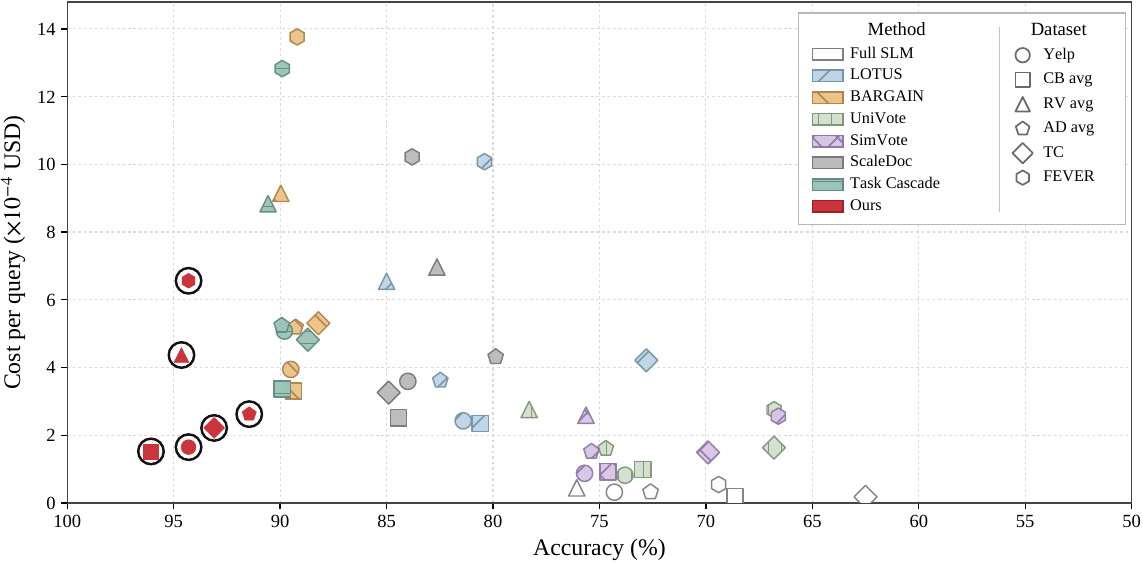}
\caption{Accuracy--cost tradeoff, each point a method's mean over the settings in a dataset group. The Accuracy axis increases to the left, so the lower-left corner (high accuracy, low cost) is best.}

\label{fig:tradeoff}
\end{figure}

\subsection{Cascade-Structure Ablation (RQ2)}
\label{sec:exp:multigranularity}
Each variant changes which decisions are available to the cascade and is described as being recalibrated at the same oracle-relative target. Figure~\ref{fig:ablation_radar}(a,b) reports the resulting benchmark Accuracy and nominal cost. Removing Stage~1 changes the full method's mean Accuracy from 0.938 to 0.923 and its cost from 2.968 to $5.570\times10^{-4}$ per predicate--record (1.88$\times$). Removing Stage~2 gives Accuracy 0.914 and cost $4.871\times10^{-4}$ (1.64$\times$). These measurements support retaining both stages for this workload and configuration. They do not prove that both stages are necessary for every predicate, or that a ground-truth Accuracy above 0.9 validates the oracle-relative certificate.

\subsection{Component Ablation (RQ3)}
\label{sec:exp:ablation}
Two components change the selected operating point: Neighbor Support (NS) and evidence retrieval. Figure~\ref{fig:ablation_radar}(c,d) reports Accuracy 0.901 without NS, versus 0.938 for the full method, and nominal cost $6.733\times10^{-4}$ versus $2.968\times10^{-4}$. Removing NS changes scores, grids, and subsequently selected thresholds, so its effect on deferral is not determined by the sign of the score adjustment alone. Removing RRF or the reranker gives Accuracy 0.934 or 0.926 and cost $3.326\times10^{-4}$ or $3.502\times10^{-4}$, respectively. These are observations for the displayed ablations, not evidence that every extraction or routing replacement improves every record.

The two search ablations in Figure~\ref{fig:ablation_radar}(a,b) are stage-wise tuning (w/o Joint Search) and flat grid-wide Bonferroni (w/o Fixed-Sequence). Their displayed Accuracy values are 0.906 and 0.908, and nominal costs are $7.866\times10^{-4}$ and $7.135\times10^{-4}$: 3.2 and 3.0 pp lower Accuracy and 2.65$\times$ and 2.40$\times$ higher cost than the full method. A flat correction tests each grid point at $\delta/(L_1L_2)$; fixed-sequence testing uses $\delta/L_1$ and stops each track after its first failed test. The larger per-test budget does not establish uniform power dominance because the stopping rules visit different candidates. The stage-wise variant's allocation and selection procedure require a separate validity check from its implementation.

\begin{figure}[!t]

\centering
\includegraphics[width=\columnwidth]{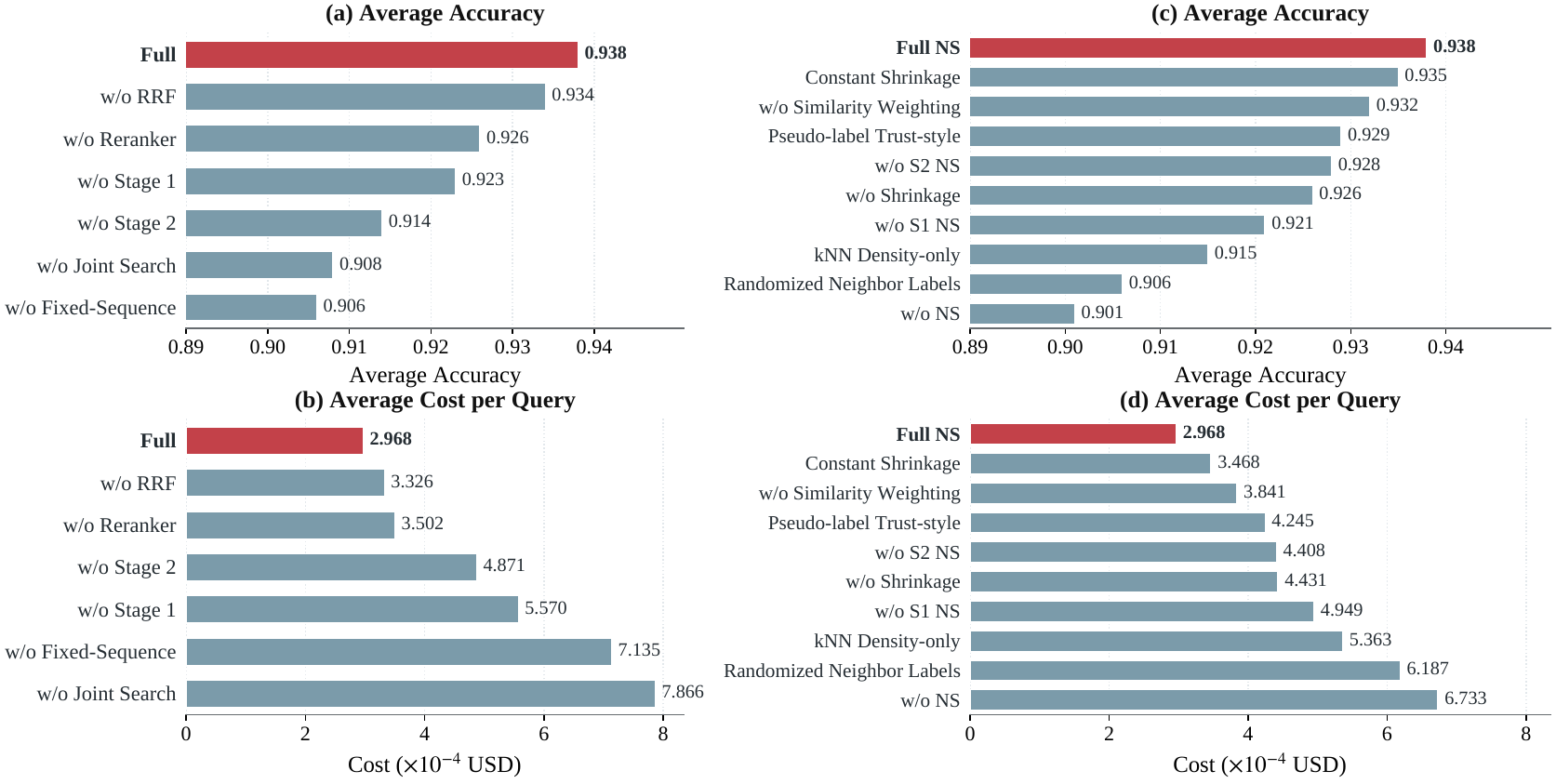}
\caption{Component-removal ablation: (a,b) pipeline and search variants; (c,d) routing-score variants. The panels report benchmark Accuracy and nominal cost; averages are equal-weight arithmetic means across settings.}

\label{fig:ablation_radar}
\end{figure}

\subsection{Calibration Diagnostics (RQ4)}
\label{sec:exp:validity}
The results above summarize operating points, but the guarantee concerns the distribution over calibration draws. We repeat calibration and deployment on 20 independent draws of $S$ per setting at the default $\tau$, $\delta$, and calibration size $n$. The intended protocol draws a fresh sample, obtains its fixed oracle labels, reruns Algorithm~\ref{alg:certify}, and evaluates the selected configuration. Re-querying a stochastic oracle would change the target population and would need separate treatment.

\noindent\textbf{Validity.}
The primary check is the frequency with which oracle-relative accuracy falls below $\tau$. The current routing table records six violations on AD-Q2 and three on AD-Q3, totaling 9/260 (3.46\%). Earlier narrative values of 3/260 and one violation per setting conflict with this table and are not used here. In particular, the reported AD-Q2 rate is 6/20; pooled results cannot establish the per-setting guarantee. Per-draw logs are needed to reconcile these counts and assess the implementation before claiming empirical validation of the certificate. Differences of 0.001 between subtracting rounded means and the signed-gap column can arise from independent rounding. They are not, by themselves, contradictions; the unrounded means and per-draw results are needed to verify that column and its uncertainty summaries.

\noindent\textbf{Power and the label floor.}
Certifying a track requires wealth $L_1/\delta=640$ at $L_1=32$ and $\delta=0.05$. The correct finite-population zero-error envelope is the product in Eq.~\eqref{eq:zero_envelope}, not a general $n\,\mathrm{KL}$ bound. For Codebase ($N=9378$), its values are 165.633 at $n=94$, 616.076 at $n=118$, and 650.788 at $n=119$. Thus 119 is a necessary zero-error feasibility floor under the stated betting cap. It does not ensure that the implemented adaptive bet reaches 640: actual bets may lie below the cap. The manuscript reports $n_0=119$ when the nominal 1\% budget is smaller. However, the released code at revision \texttt{99adc74} yields wealth 463.601 on 119 zero errors and first reaches 640 at 125 zero errors. Its split script also defaults to $\min(N,\max(500,\lceil0.01N\rceil))$. These are different rules; the reported runs cannot be identified with that revision without their manifests and wealth traces. Appendix~\ref{app:power} gives the derivation and a counterexample to the previous KL claim.

\subsection{Calibration Sensitivity (RQ5)}
\label{sec:exp:sweep}
Figure~\ref{fig:hyperparam_sensitivity} varies the accuracy target $\tau$, failure budget $\delta$, and calibration size $n$. These parameters play different roles: $\tau$ determines the admissible error rate, $\delta$ determines the strength of evidence required for certification, and $n$ determines how many oracle labels are available to provide that evidence.

\noindent\textbf{Accuracy target.}
The reported operating points generally move toward higher accuracy and higher cost as $\tau$ increases. A stricter target makes it harder to accept small-model predictions, so more records may reach the oracle. This is a tendency of the selected configurations rather than a claim that every individual routing threshold changes monotonically. Ground-truth accuracy in the plot is a measured outcome; the theorem concerns oracle-relative accuracy.

\noindent\textbf{Failure budget.}
Across the reported range $\delta\in[0.005,0.10]$, the operating points change less than in the accuracy-target sweep. The test threshold depends on $\log(L_1/\delta)$, while selected plans change only when a candidate enters or leaves the certified set. A range of failure budgets can therefore select the same plan. This observation does not establish that $\delta$ is unimportant or that the certificate has been empirically validated.

\noindent\textbf{Calibration size.}
A larger sample provides more evidence for certification, but requires additional oracle labels. Consequently, a lower deployment fallback rate need not imply a lower total cost. The reported sweep illustrates this tradeoff; any end-to-end comparison must count calibration labels and cached predictions once, using the accounting in Appendix~\ref{app:cost}. We also examine how replacing the small model, embedding model, and reranker changes the selected operating point (Section~\ref{sec:exp:backbones}).

\begin{figure}[!b]

\centering
\includegraphics[width=\columnwidth]{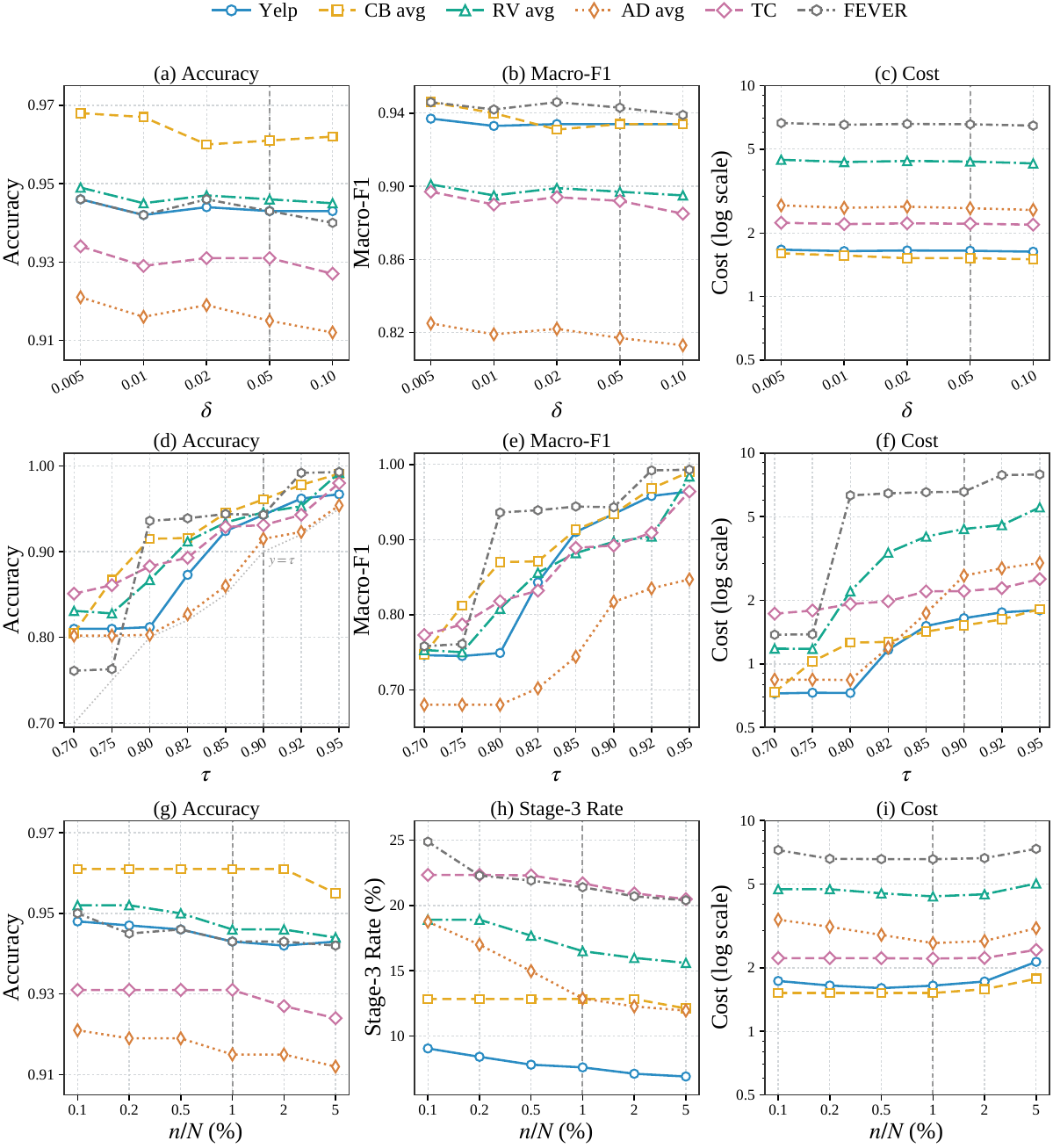}
\caption{Sensitivity to the target failure probability $\delta$,
target accuracy $\tau$, and calibration ratio $n$.}

\label{fig:hyperparam_sensitivity}
\end{figure}

\subsection{Model-Backbone Robustness}
\label{sec:exp:backbones}
\noindent\textbf{Measured model-replacement behavior.}
Table~\ref{tab:model_sensitivity} replaces the SLM, embedding, and reranking backbones while holding the cascade and operating point ($\tau{=}0.9$, $\delta{=}0.05$, and $n$ under the default calibration rule) fixed, re-certifying each replacement. Every backbone achieves measured ground-truth Accuracy of at least 0.901, with the lowest value observed for Qwen3/0.6B and Llama-3.2/1B on RV-Q1. The maximum within-dataset Accuracy spread is 6.1 percentage points.

The main difference is in Stage~3 routing. On CB-Q2, Qwen3/0.6B routes 54.1\% of records to the oracle, compared with 6.1\% for Qwen3/4B. Replacing the default embedding backbone increases the CB-Q2 fallback rate from 6.1\% to 12.1--13.8\%. The trend is capability-dependent but not strictly monotone: Qwen3/8B reduces fallback on Yelp (6.2\% vs.\ 7.6\%) and FEVER (10.3\% vs.\ 21.4\%), while increasing it on CB-Q2 (12.7\% vs.\ 6.1\%) and RV-Q1 (7.8\% vs.\ 3.8\%). Changes in prediction confidence and the resulting quantile grid influence the selected routing thresholds. The theorem remains applicable after recalibration only when the fixed-output and sampling assumptions hold; the displayed ground-truth metrics do not check those assumptions. Overall Accuracy can nevertheless mask differences in Macro-F1: on Yelp, replacing Qwen3/4B with Llama-3.2/1B reduces Macro-F1 from 0.934 to 0.888 while Accuracy changes from 0.943 to 0.902.

\begin{table}[!t]

  \centering
  \caption{Model replacement study. A/F denotes Accuracy/Macro-F1; S3 denotes the Stage~3 fallback rate. Bold marks the best A/F and the lowest S3 within each type. BGE-M3 in the Rer. block is a bi-encoder similarity control without cross-encoder reranking.}
  \label{tab:model_sensitivity}
  \scriptsize
  \setlength{\tabcolsep}{2.0pt}
  \renewcommand{\arraystretch}{0.88}
  \resizebox{\columnwidth}{!}{%
  \begin{tabular}{@{}ll*{4}{cc}@{}}
  \toprule
  \textbf{Type} & \textbf{Model}
  & \multicolumn{2}{c}{\textbf{Yelp}}
  & \multicolumn{2}{c}{\textbf{CB-Q2}}
  & \multicolumn{2}{c}{\textbf{FEVER}}
  & \multicolumn{2}{c}{\textbf{RV-Q1}} \\
  \cmidrule(lr){3-4}
  \cmidrule(lr){5-6}
  \cmidrule(lr){7-8}
  \cmidrule(lr){9-10}
  & & \textbf{A/F} & \textbf{S3(\%)}
  & \textbf{A/F} & \textbf{S3(\%)}
  & \textbf{A/F} & \textbf{S3(\%)}
  & \textbf{A/F} & \textbf{S3(\%)} \\
  \midrule
  SLM & Qwen3/0.6B~\cite{qwen3_2025} & 0.904/0.890 & 46.8 & 0.904/0.901 & 54.1 & 0.903/0.902 & 34.2 & 0.901/0.901 & 32.7 \\
  SLM & Qwen3/1.7B~\cite{qwen3_2025} & 0.916/0.903 & 30.2 & 0.917/0.914 & 36.8 & 0.912/0.912 & 17.4 & 0.924/0.924 & 12.6 \\
  SLM & Qwen3/4B~\cite{qwen3_2025} & \textbf{0.943/0.934} & 7.6 & 0.956/0.956 & \textbf{6.1} & \textbf{0.943/0.943} & 21.4 & 0.955/0.955 & \textbf{3.8} \\
  SLM & Qwen3/8B~\cite{qwen3_2025} & 0.939/0.932 & \textbf{6.2} & \textbf{0.958/0.957} & 12.7 & 0.941/0.941 & \textbf{10.3} & \textbf{0.958/0.958} & 7.8 \\
  SLM & Llama-3.2/1B~\cite{meta2024llama32} & 0.902/0.888 & 49.6 & 0.903/0.899 & 60.9 & 0.902/0.901 & 42.1 & 0.901/0.901 & 29.4 \\
  SLM & Llama-3.2/3B~\cite{meta2024llama32} & 0.909/0.895 & 39.7 & 0.909/0.906 & 45.4 & 0.907/0.906 & 31.8 & 0.905/0.905 & 28.9 \\
  \midrule
  Emb. & Qwen3/4B & \textbf{0.943/0.934} & \textbf{7.6} & \textbf{0.956/0.956} & \textbf{6.1} & \textbf{0.943/0.943} & \textbf{21.4} & \textbf{0.955/0.955} & \textbf{3.8} \\
  Emb. & Qwen3/0.6B & 0.921/0.910 & 11.6 & 0.930/0.928 & 13.8 & 0.922/0.922 & 23.4 & 0.919/0.919 & 5.9 \\
  Emb. & BGE-M3~\cite{chen2024m3} & 0.918/0.907 & 10.9 & 0.939/0.937 & 12.9 & 0.925/0.925 & 22.6 & 0.913/0.913 & 5.2 \\
  Emb. & E5-L~\cite{wang2022e5} & 0.924/0.913 & 11.3 & 0.935/0.933 & 12.1 & 0.922/0.922 & 23.2 & 0.924/0.924 & 6.6 \\
  \midrule
  Rer. & Qwen3/0.6B & 0.943/0.934 & \textbf{7.6} & 0.956/0.956 & \textbf{6.1} & 0.943/0.943 & \textbf{21.4} & 0.955/0.955 & \textbf{3.8} \\
  Rer. & Qwen3/4B & \textbf{0.948/0.941} & 9.1 & \textbf{0.961/0.960} & 8.5 & \textbf{0.951/0.951} & 24.1 & \textbf{0.962/0.962} & 4.2 \\
  Rer. & BGE-M3~\cite{chen2024m3} & 0.925/0.915 & 11.2 & 0.930/0.927 & 13.3 & 0.928/0.928 & 25.4 & 0.915/0.915 & 4.7 \\
  \bottomrule
  \end{tabular}%
  }
\end{table}

\subsection{Threats to Validity}
\label{sec:exp:threats}
\noindent\textbf{Oracle as ground-truth proxy.}
The certificate is, strictly, oracle-relative: the theorems bound $\mathbb P(\mathrm{Acc}_{\ML}(\hat\phi)\geq\tau)\geq1-\delta$, and transfer to $f^*$ uses Eq.~\eqref{eq:gt_bridge}, which needs an upper bound on oracle--ground-truth disagreement while the absolute signed accuracy gap supplies only a lower bound (Section~\ref{sec:exp:validity}); we clear the GT target by measurement rather than by proof. The current experiment certifies accuracy against $\ML$, not $f^*$. Changing calibration labels to ground truth changes the loss at every stage, including oracle fallback. A mean-loss test can still be constructed, but an end-to-end guarantee requires a safe or separately certified fallback, or a no-decision outcome. It does not follow from changing the label source or multiplicity correction alone (Appendix~\ref{rem:bounded}).

\noindent\textbf{Benchmark and cost accounting.}
The Yelp ground truth keeps only records on which three labelers agree, which removes ambiguous cases, making the benchmark easier for every method and affecting the observed oracle-relative gap there: the absolute Yelp numbers and the relative ranking may therefore differ on an unfiltered population. The local-compute portion requires a documented resource and pricing ledger before it can be repriced. Because we compare systems at their delivered operating points rather than sweeping each baseline's own knob, the plotted cost gaps are not matched-accuracy comparisons or a frontier comparison. The guarantee is over the finite population $\mathcal I$, so each deployment draws a fresh sample; cross-corpus transfer without recalibration is not covered.

\noindent\textbf{Certification conservatism and operating point.}
The single-point test is one-sided, so a feasible configuration can fail certification and be dropped from $\mathcal C$; \system{} returns the cheapest certified candidate, not the global optimum (Section~\ref{sec:guarantee:scope}). The reported ground-truth Accuracy of \system{} meets 0.9 in every setting, and Figure~\ref{fig:tradeoff} summarizes the measured accuracy--cost operating points. Finally, the nominal token costs do not establish runtime, throughput, GPU-hours, or latency savings. Batching changes the invocation pattern; its wall-clock effect requires measurement. Batch composition is fixed by document order before calibration, so it does not interact with the sampling argument, but a throughput-oriented deployment would want the wall-clock accounting we omit.
The violation-count discrepancy and cached-cost accounting still require verification against run logs; the present results do not establish empirical validity of the certificate.

}

% =============================================================
% SECTION 6: RELATED WORK
% =============================================================
{
\section{Related Work}
\label{sec:related}

\noindent\textbf{Semantic filtering and query approximation.}
Approximate predicate evaluation predates LLMs~\cite{lu2018probabilisticpredicates}; recent systems expose LLM-powered operations as declarative primitives: LOTUS~\cite{patel2024lotus} introduces semantic operators with statistical guarantees on expected loss against a gold algorithm; Palimpzest~\cite{liu2025palimpzest}, DocETL~\cite{shankar2024docetl}, and MOAR~\cite{wei2025moar} optimize analytical pipelines; SUQL~\cite{liu2024suql}, ZenDB~\cite{lin2024zendb}, and UQE~\cite{dai2024uqe} query structured and unstructured data; and Abacus~\cite{russo2025abacus}, TAG~\cite{biswal2024tag}, and TASTI~\cite{kang2022tasti} add cost-based optimization, an operator interface, and semantic indexes. Closer to our cost focus, UniVote/SimVote~\cite{hou2026beyondlinear}, Fast Semantic Filtering~\cite{kim2026fastsemanticfiltering}, and AI Query Approximation~\cite{chung2026aiqueryapprox} cut LLM calls under an accuracy target, and TRAQ~\cite{li2024traq} guarantees answer-set coverage; these works provide related approaches to reducing semantic execution cost.

\noindent\textbf{Model routing and cascades.}
FrugalGPT~\cite{chen2023frugalgpt}, RouteLLM~\cite{ong2024routellm}, AutoMix~\cite{aggarwal2024automix}, Hybrid LLM~\cite{ding2024hybridllm}, SMART~\cite{jo2024smart}, ThriftLLM~\cite{huang2025thriftllm}, and Universal Model Routing~\cite{jitkrittum2025uniroute} study cost-aware model selection; C3PO~\cite{valkanas2025c3po} and UCCI~\cite{kotte2026ucci} add probabilistic cost constraints and calibrated uncertainty to cascade routing; SUPG~\cite{kang2020supg} gives proxy-based selection with precision/recall guarantees; Streaming~\cite{liskowski2026streamingcascades} and Task Cascades~\cite{shankar2026taskcascades} study streaming and generalized cascades. BARGAIN~\cite{10.1145/3769776} is closely related: it selects a routing threshold for a cheap-model/oracle cascade and provides task-specific quality guarantees. Our formulation fixes an evidence/full-document/oracle structure and compares pairs of thresholds using per-record token costs. This is a difference in the optimization studied here, rather than a claim that prior cascades cannot account for token prices or accept alternative input representations.

\noindent\textbf{Finite-population calibration.}
Our guarantee builds on the Waudby-Smith and Ramdas (WS\&R) betting framework for bounded means~\cite{waudbysmith2024}. Conformal Risk Control (CRC)~\cite{angelopoulos2024crc} controls expected loss. Selective CRC~\cite{xu2025selectivecrc} combines sample selection with risk control, including a calibration-only variant with probably approximately correct (PAC) guarantees, while Adaptive Conformal Inference~\cite{gibbs2021adaptive} targets online distribution shift. The closest methodology is Learn-then-Test (LTT)~\cite{angelopoulos2021learntest}, which, like us, selects configurations from a grid under family-wise error control; we differ in the sampling model and in what is being selected. LTT provides a general framework for calibrating collections of configurations and is not limited to scalar rules. We instantiate related multiple-testing tools for a fixed finite deployment population sampled without replacement, using residual-population centering for finite-sample validity (Section~\ref{sec:guarantee:wsr}). Our candidates are coupled threshold pairs whose Stage-2 populations depend on $\lambda_1$ (Section~\ref{sec:guarantee:multi}); the test can be conservative and does not estimate error exactly.

Task Cascades already combines varying document portions with cascade-level accuracy certification. Its guarantee adjustment follows a prescribed threshold sequence after assembly; our fixed three-stage design instead compares certified threshold pairs by token-aware cost. Thus neither shorter inputs nor statistical certification alone distinguishes our contribution.

}

% =============================================================
% SECTION 7: CONCLUSION
% =============================================================
{
\section{Conclusion}
\label{sec:conclusion}
We presented \system{}, which selects two thresholds for a fixed three-stage semantic filter to reduce modeled execution cost under an oracle-relative error constraint. The optimization is generally nonconvex, but its error is monotone along each fixed-first-threshold track. Our method uses this structure to certify candidates from a finite calibration sample and selects the cheapest certified pair, without claiming recovery of the true feasible optimum. Experiments cover six datasets and 13 settings; the main table reports the highest benchmark accuracy and Macro-F1 in 12 settings each. The reported violation counts and end-to-end cost accounting still require reconciliation. Future work includes other semantic operators and recalibration under distribution shift.

}

% References are outside the 12-page main text.
\clearpage
% Reference 32 is in the first column of the final bibliography page; recheck after edits.
\IEEEtriggeratref{32}
\IEEEtriggercmd{\balance}
\bibliographystyle{IEEEtran}
\bibliography{references}

\clearpage
\nobalance
{
\appendices
\section{Technical Report: Proofs and Supplementary Details}
\label{app:technical-report}
This appendix contains the technical-report material referenced in the main text.

\subsection{Proof of Theorem~\ref{thm:coverage}}
\label{app:proof:coverage}
\begin{proof}
Fix the entire population, its potential model outputs, and all candidate-specific test rules independently of the sample draw. Write $\theta=N^{-1}\sum_i e_i(\phi)$ and use the filtration and bets in Appendix~\ref{app:betting}. Uniform ordered sampling gives Eq.~\eqref{eq:residual_mean}; in particular its right-hand side is in $[0,1]$ on every history of positive probability.

For $u\leq\theta$, the unclipped quantity $(Nu-S_{t-1})/(N-t+1)$ is no larger than $\mu_t=\mathbb E[X_t\mid\mathcal F_{t-1}]$. Clipping to $[0,1]$ preserves $q_t(u)\leq\mu_t$. For $q_t<1$, the admissible bet ensures $1+\kappa_t(q_t-X_t)\geq1-\kappa_t(1-q_t)\geq1/2$. For $q_t=1$, any finite nonnegative bet gives a factor at least one; under the null, $\mu_t=1$ and $X_t=1$ almost surely, so the factor equals one. No infinite bet is permitted.

Since $\kappa_t$ and $q_t$ are measurable before observing $X_t$,
\[
\begin{aligned}
&\mathbb E[M_t^+(u)\mid\mathcal F_{t-1}]\\
&\quad=M_{t-1}^+(u)\{1+\kappa_t(u)(q_t(u)-\mu_t)\}\\
&\quad\leq M_{t-1}^+(u).
\end{aligned}
\]
Thus $M_t^+(u)$ is a nonnegative supermartingale with initial value one. At the fixed sample size, $\mathbb E[M_n^+(u)]\leq1$. Markov's inequality at $u=\varepsilon<\theta$ yields $\Pr(M_n^+(\varepsilon)\geq1/\beta)\leq\beta$. This proves the stated fixed-time bound; Ville's inequality also bounds crossing at any time. If the fixed candidate family is itself random independently of the sampling permutation, the same argument holds conditionally on that family, and averaging preserves the bound.
\end{proof}

\subsection{Proof of Lemma~\ref{lem:monotone}}
\label{app:proof:monotone}
\begin{proof}
The scores $\tilde{z}_i^{(2)}$ are frozen before calibration, so the accept
indicator is nested, $A_i^{(2)}(\lambda_{2,m})\leq A_i^{(2)}(\lambda_{2,m'})$ for $m<m'$.
With $\lambda_1$ fixed, the first term of~\eqref{eq:error_indicator} is unchanged and the second equals
$g_i A_i^{(2)}(\lambda_2)$ with the $\lambda_2$-free factor $g_i=\mathbf{1}[i\in\mathcal{I}_2^{\max}]D_i^{(1)}\mathbf{1}[\hat{y}_i^{(2)}\neq y_i]\geq0$,
so each $e_i$ is non-decreasing in $m$.
\end{proof}

\subsection{Proof of Lemma~\ref{lem:certified_set}}
\label{app:proof:certified_set}
\begin{proof}
Condition on the fixed population, candidate family, orders, and test rules. On track $k$, let $m_k^\star=\min\{m:E_{k,m}>\varepsilon\}$, if that set is nonempty. This index is fixed with respect to calibration sampling, even though its value is unknown to the algorithm. If there is no such index, that track cannot certify an infeasible candidate.

Otherwise, certifying any infeasible candidate requires certifying the first infeasible candidate $m_k^\star$: every earlier candidate must pass because the scan stops at its first failure. Therefore
\[
\begin{gathered}
\{\text{track }k\text{ certifies an infeasible pair}\}\\
\subseteq\{M_n^+(\varepsilon;\phi_{k,m_k^\star})\geq1/\eta_f\}.
\end{gathered}
\]
The latter event has probability at most $\eta_f$ by Theorem~\ref{thm:coverage}. Summing over the $L_1$ tracks gives at most $L_1\eta_f=\delta$. Independence between tests or tracks is not needed, so all candidates may reuse the same sample. Monotonicity is not needed for this first-error inclusion; it explains why a stringent-to-lenient order is useful for threshold search. Reordering or restarting tests after observing calibration labels is outside this proof.
\end{proof}

\subsection{Proof of Theorem~\ref{thm:guarantee}}
\label{app:proof:guarantee}
\begin{proof}
Let $\mathcal{G}=\{\forall\phi\in\mathcal{C}:E(\phi)\leq\varepsilon\}$, so $\mathbb{P}(\mathcal{G})\geq1-\delta$ by Lemma~\ref{lem:certified_set} under the theorem assumptions.
On $\mathcal{G}$, either $\mathcal{C}\neq\varnothing$ and $\hat{\phi}\in\mathcal{C}$ gives $E(\hat{\phi})\leq\varepsilon$, or the fallback
emits $y_i$ on every record and $E(\hat{\phi})=0$; either way the target holds,
so $\mathbb{P}(\mathrm{Acc}_{\ML}(\hat{\phi})<\tau)\leq\mathbb{P}(\mathcal{G}^c)\leq\delta$. Because $\mathcal{G}$ controls \emph{every}
member of $\mathcal{C}$ simultaneously, validity does not depend on which
member the label-free cost rule selects, and the $\arg\min$ over the
finite $\mathcal{C}$ is attained.
\end{proof}

\subsection{Proof of Theorem~\ref{thm:complexity}}
\label{app:proof:complexity}
\begin{proof}
There are at most $N$ routing scores at each SLM stage, so sorting them and selecting grid quantiles costs $O(N\log N+G)$ time, including grid materialization. A direct scan evaluates each candidate's routing decisions and sums its cost contributions in $O(N)$ time, giving $O(GN)$ for all candidates. Each candidate's certificate uses at most $n$ constant-time betting updates, and at most $G$ candidates are tested. Early stopping within tracks can reduce this work but does not increase its $O(Gn)$ worst-case bound. Selecting the cheapest certified candidate costs $O(G)$, and producing final routing decisions costs $O(N)$. Cached scalar scores, predictions, and labels occupy $O(N)$ space; candidate costs and certification results occupy $O(G)$. Processing one candidate at a time avoids storing an $N\times G$ error matrix. Uniform ordered sampling can be implemented with a partial permutation in $O(N)$ time and space. The nonempty-grid condition makes $O(N+G)$ additive work dominated by the stated time bound (including $N=1$). Finally, calibration labels are queried for $S$, and deployment queries only the uncached members of $R$, yielding the stated oracle count; for all-oracle fallback take $R=\mathcal I$. The bound counts real-arithmetic operations and excludes arbitrary-precision bit complexity. An empty grid can instead return the fallback immediately. If the Stage~2 pool is empty, a singleton dummy second-stage grid suffices, since all candidates stop at Stage~1.
\end{proof}

\subsection{Proof of Theorem~\ref{thm:threshold_structure}}
\label{app:proof:threshold_structure}
\begin{proof}
Routing indicators change only when a threshold crosses a score, proving the piecewise-constant property. For nonconvex cost, take one record with $\tilde z^{(1)}=0.5$, set $c_2=1$, and fix $\lambda_2$ so Stage~2 accepts. After subtracting $C_1$, costs at $\lambda_1=0.4,0.8$ are $0,1$, whereas their midpoint $0.6$ has cost $1>1/2$, violating convexity.

For a nonconvex feasible set, take one record with both scores $0.5$, $a=0$, $b=1$, and $\varepsilon=0.1$. The pair $(0.4,0.2)$ accepts the correct Stage~1 answer; $(0.8,0.6)$ uses the oracle. Both have zero error. Their midpoint $(0.6,0.4)$ accepts the incorrect Stage~2 answer and has error one, so is infeasible.

Increasing $\lambda_1$ changes a record's cost from $c_{1i}$ to $c_{1i}+c_{2i}+D_i^{(2)}c_{Li}$; increasing $\lambda_2$ can only add oracle cost. Both changes are nonnegative. For fixed $\lambda_1$, increasing $\lambda_2$ decreases $1-D_i^{(2)}$, whose coefficient $D_i^{(1)}b_i$ in Eq.~\eqref{eq:population_error} is nonnegative, so error cannot increase. In contrast, crossing a Stage~1 score changes error by $(1-D_i^{(2)})b_i-a_i$, which can have either sign. For an explicit example, take two records with Stage~1 scores $0.3,0.7$, Stage~2 scores both $0.8$, $a=(0,1)$, $b=(1,0)$, and $\lambda_2=0.5$. Their mean errors at $\lambda_1=0.2,0.5,0.8$ are $1/2,1,1/2$, respectively. Thus error need not be monotone in the first threshold.
\end{proof}

\subsection{Evidence Extraction and Stage Details}
\label{app:implementation}

\subsubsection{Evidence locality}
\label{sec:system:stage1:locality}
The chunk view reduces input length when a short passage contains sufficient evidence. Whether it improves a prediction is data-dependent; it is evaluated separately from the full-document view.

\subsubsection{Adaptive Evidence Extraction}
\label{sec:system:stage1:extraction}
\label{sec:system:stage1:context}
Predicates rely on different signals and relevant content varies in
density across records, so a fixed top-$k$ rule does not fit.
\system{} sizes the evidence set per record with a two-cutoff retrieval pipeline.
The document $d_i$ is split into sentences (BlingFire) and formed into
overlapping three-sentence windows (stride $1$).
Candidates are ranked by reciprocal rank fusion (RRF)~\cite{cormack2009rrf} of
BM25 and dense (Qwen3-4B-Emb~\cite{qwen3embedding_2025,qwen2025qwen3_embedding_4b_modelcard}) scores,
$\mathrm{RRF}(c)=\tfrac{1}{k_0+\mathrm{rank}_{\mathrm{BM25}}(c)}
+\tfrac{1}{k_0+\mathrm{rank}_{\mathrm{emb}}(c)}$ ($k_0{=}60$), combining
exact lexical hits with semantic matches.
A first, recall-favoring cutoff keeps candidates up to the largest gap in
the sorted RRF scores, plus a buffer of five; a cross-encoder reranker (Qwen3-0.6B-Reranker~\cite{qwen3embedding_2025,qwen2025qwen3_reranker_0p6b_modelcard}) rescores the survivors, and a second, precision-favoring cutoff (no buffer, capped
at $5$ chunks) keeps them up to the largest reranker-score gap.
Selected chunks are merged in document order, overlapping sentences deduplicated, into $e_{\mathrm{chunk}}(d_i)$.

\subsubsection{Batched Inference and Confidence Scoring}
\label{sec:system:stage1:routing}
\system{} packs multiple query--evidence pairs into one batched context
with explicit DocID delimiters,
\[
C_B=\mathrm{concat}[\langle\mathrm{DocID}{=}i,\,q_i,\,e_{\mathrm{chunk}}(d_i)\rangle]_{i\in B},
\]
where a token budget $B_{\max}$ caps batch size.
The SLM emits one binary label per document identifier (DocID) in a single pass; any malformed
output is marked \emph{defer}, which can only shrink the accept set, preserving
the guarantee. For each record, \system{} reads the log-probability margin at its label position,
$\ell_i=\log p(\texttt{yes}\mid C_B,\mathrm{prefix}_i)
-\log p(\texttt{no}\mid C_B,\mathrm{prefix}_i)$, predicts
$\hat{y}_i^{(1)}=\mathbf{1}[\ell_i>0]$, and, since the margin scale shifts with batch
length, maps it to a base confidence \(b_i^{(1)}\allowbreak=\allowbreak\sigma\!\left(\ell_i/T_{\mathrm{temp}}\right)\allowbreak\in[0,1]\)
($T_{\mathrm{temp}}{=}15$), comparable across batches; logit confidence is preferred over
verbalized self-assessment, poorly calibrated for
routing~\cite{jin2024llmuncertainty}. This base score is fused with neighborhood evidence
into the adaptive routing score $\tilde{z}_i^{(1)}$
(Section~\ref{sec:system:neighbor_support}), and Stage~1 accepts when
$\tilde{z}_i^{(1)}\geq\lambda_1$, with $\lambda_1$ selected by certification.

\subsubsection{Stage~2: Full-Document Filtering}
\label{sec:system:doc}
Records that any candidate $\lambda_1$ could defer form the maximal
Stage-2 subset
\begin{equation}
 \mathcal{I}_2^{\max} = \{i \in \mathcal{I} : \tilde{z}_i^{(1)} <
 \lambda_{1,\max}\}, \qquad \lambda_{1,\max} = \max \Lambda_1,
 \label{eq:i2max}
\end{equation}
which depends only on the fixed $\lambda_{1,\max}$, not on any
candidate $\lambda_1$, so running Stage~2 on $\mathcal{I}_2^{\max}$ covers every record
that could reach it. This full pass is a one-time \emph{search} computation supplying
the Stage-2 scores for each candidate's label-free routing counts, not deployment
routing, since under $\hat{\phi}$ a record accepted at Stage~1 is never sent to Stage~2;
it is a constant offset across candidates, counted in full in the reported cost
(Section~\ref{sec:guarantee:scope}).
Stage~2 re-evaluates each record individually, without batch packing, removing the
Stage~1 batch error modes (cross-document attention, DocID--label misalignment, evidence
truncation), and accepts when $\tilde z_i^{(2)}\geq\lambda_2$, with the score recomputed from full-document embeddings.

\subsubsection{Stage~3: Oracle Fallback}
\label{sec:system:fallback}
The residual set (records deferred by both SLM stages under $\hat{\phi}$) holds the
records that require the oracle. \system{} submits each to $\ML$ with the full document
and predicate, returning the binary label and a short justification for auditability.
The Stage~3 rate reflects predicate difficulty without tuning, from $2.5\%$
on CB-Q1 to $31.2\%$ on RV-Q3 (Table~\ref{tab:routing}).
Because Stage~3 records are scored against the oracle label $y_i=\ML(q_i,d_i)$,
they contribute zero loss to the certification target by construction; the tier is
warranted where SLM capacity runs out, as on CB-Q3 where the full-SLM reaches only
$61.4\%$ accuracy (Section~\ref{sec:exp:main}).

\subsection{Routing-Score Equations}
\label{app:score}

\noindent\textbf{The idea.}
Two records can share the same base confidence $b_i^{(s)}$ while only one deserves acceptance (Figure~\ref{fig:motivation}c), so we want each record's effective threshold to track its difficulty, with no per-record tuning and no held-out labels. A free second cue makes this possible: a record whose embedding neighbors received the same prediction is likelier correct than one that disagrees, so \system{} nudges the base confidence by how strongly a record's neighbors agree with its prediction.

\noindent\textbf{Neighbor support.}
Let $v_i^{(s)}$ be the unit-normalized Stage-$s$ embedding and
$\mathcal{N}_K^{(s)}(i)$ its top-$K$ cosine neighbors within the stage's own pool
($\mathcal{I}$ at Stage~1, $\mathcal{I}_2^{\max}$ at Stage~2), with weights
$w_{ij}^{(s)}=\max\{\langle v_i^{(s)},v_j^{(s)}\rangle,0\}$ and total mass
$W_i^{(s)}=\sum_j w_{ij}^{(s)}$. \emph{Neighbor support}
$NS_i^{(s)}\in[0,1]$ is the similarity-weighted fraction of those neighbors
whose prediction matches $i$'s own $\hat{y}_i^{(s)}$ (set to $\tfrac12$ when
$W_i^{(s)}=0$); it reads neighbor \emph{predictions}, never labels.

\noindent\textbf{Shrinkage.}
A neighborhood with few or distant neighbors carries little evidence, so we
trust $NS_i^{(s)}$ only in proportion to that evidence, shrinking it toward
the uninformative value $\tfrac12$:
\begin{equation}
 \begin{gathered}
 \widetilde{NS}_i^{(s)} = \rho_i^{(s)}\,NS_i^{(s)}
 + \big(1-\rho_i^{(s)}\big)\tfrac12,
 \\
 \rho_i^{(s)} = \frac{W_i^{(s)}}{W_i^{(s)}+\sigma_0^2}\in[0,1].
 \end{gathered}
 \label{eq:theta}
\end{equation}
An evidence-rich neighborhood keeps
$\widetilde{NS}_i^{(s)}\!\approx\!NS_i^{(s)}$, an evidence-poor one returns it
to $\tfrac12$. Fixing $\sigma_0^2$ to the median of $W_i^{(s)}$
over that pool needs no tuned constant and makes
$\widetilde{NS}_i^{(s)}$ invariant to global rescaling of the similarities.

\noindent\textbf{Score.}
The routing score adds a bounded neighbor adjustment to the base confidence
in log-odds space,
\begin{equation}
 \tilde{z}_i^{(s)} =
 \operatorname{lgt}\!\big(b_i^{(s)}\big)
 + \gamma_s\,\big(2\widetilde{NS}_i^{(s)}-1\big),
 \label{eq:adaptive_score}
\end{equation}
where $\operatorname{lgt}(b)=\log\frac{b+\varepsilon_\ell}
{1-b+\varepsilon_\ell}$ is a smoothed log-odds link
($\varepsilon_\ell{=}10^{-3}$) and the neighbor term
$2\widetilde{NS}_i^{(s)}-1\in[-1,1]$ is positive under agreement, negative
under dissent, and zero when uninformative; $\gamma_s\geq0$ (default $1$;
$\gamma_s{=}0$ recovers the pure base score) sets its weight.

\noindent\textbf{Interface to the guarantee.}
Because $\widetilde{NS}_i^{(s)}$ depends on the weights and predictions but not
on $\lambda$, each $\tilde{z}_i^{(s)}$ is frozen as $\lambda_s$ varies, so
one grid threshold acts per-record while the accept set stays nested in
$\lambda_s$. Certification needs only that the score be \emph{admissible}: computed
independently of the calibration draw and its labels, scalar, and applied through a single threshold
$A_i^{(s)}(\lambda_s)=\mathbf{1}[\tilde{z}_i^{(s)}\geq\lambda_s]$, which the default score
meets (Section~\ref{sec:guarantee}).

\subsection{Single-Candidate Betting Test}
\label{app:betting}
Condition on a fixed finite population of records and potential outputs. Decoding, batching, retrieval, score construction, candidate grids, and test specifications must be fixed independently of the calibration draw. In particular, seeing the sampled indices before choosing the family is not permitted, even if their labels have not yet been revealed. Temperature zero alone is not a mathematical guarantee of deterministic API responses; the oracle target must be a fixed function, or fixed cached potential labels generated independently of the draw.

Let $\pi_1,\ldots,\pi_n$ be the first $n$ entries of a uniform random permutation of $\mathcal I$, with fixed $1\leq n\leq N$. Preserve this order, or uniformly reshuffle the selected sample independently; do not sort it by record identity, confidence, or observed labels. For a fixed candidate define $X_t=e_{\pi_t}(\phi)$, $S_{t-1}=\sum_{k<t}X_k$, and $\mathcal F_t=\sigma(\pi_1,X_1,\ldots,\pi_t,X_t)$, including the fixed background information. Since the next index is uniform among the remaining indices,
\begin{equation}
\mathbb E[X_t\mid\mathcal F_{t-1}]
=\frac{N\theta-S_{t-1}}{N-t+1},\qquad
\theta=\frac1N\sum_i e_i(\phi).
\label{eq:residual_mean}
\end{equation}
For a target $u\in(0,1)$, set $q_t(u)=\mathrm{clip}_{[0,1]}\{(Nu-S_{t-1})/(N-t+1)\}$ and
\begin{equation}
 M_t^+(u)=\prod_{k=1}^t[1+\kappa_k(u)(q_k(u)-X_k)],
 \quad M_0^+(u)=1.
 \label{eq:wealth}
\end{equation}
Every bet must be finite, nonnegative, and $\mathcal F_{t-1}$-measurable. Require $\kappa_t(u)(1-q_t(u))\leq1/2$; when $q_t=1$, require a finite bet explicitly rather than evaluating $1/[2(1-q_t)]$. A proposed predictable nonnegative raw bet $\widetilde\kappa_t$ can be clipped to $\min\{\widetilde\kappa_t,1/[2(1-q_t)]\}$ for $q_t<1$, and left finite for $q_t=1$. The original experimental description names approximate growth-rate adaptive (aGRAPA) betting~\cite{waudbysmith2024}; the theorem covers any rule satisfying these conditions, but its power and sample-size behavior depend on the actual rule. The audited released implementation (revision \texttt{99adc74}) uses $m_t=(1/2+S_{t-1})/t$, $v_t=(1/4+\sum_{j<t}(X_j-m_j)^2)/t$, and raw bet $\widetilde\kappa_t=\max\{0,(q_t-m_t)/[v_t+(q_t-m_t)^2]\}$. These quantities use only preceding outcomes, and $v_t>0$ keeps the raw bet finite even at $q_t=1$. This identifies the current code, not necessarily the implementation used for the reported experiments. Tuning bets using later calibration outcomes is not predictable and is not covered.

Certify a candidate at level $\beta$ if $M_n^+(\varepsilon)\geq1/\beta$. This certifies the candidate through a fixed-time rejection of an excessive-error null; it does not estimate its error exactly. All proof statements are conditional on the fixed setup and therefore remain valid after integrating over independent preprocessing randomness.

\subsubsection{Zero-error feasibility envelope}
\label{app:power}
For an all-zero sample, $q_t(u)=Nu/(N-t+1)$ as long as $t<N(1-u)+1$. Maximizing each factor over the allowed bet interval gives
\begin{equation}
 U_n(u)=\prod_{t=1}^n\left(1+\frac{q_t(u)}{2(1-q_t(u))}\right),
 \qquad M_n^+(u)\leq U_n(u).
 \label{eq:zero_envelope}
\end{equation}
The bound follows because a zero-error factor $1+\kappa_t q_t$ is nondecreasing in the nonnegative bet. It is attained on this path by using the clipping cap at every step, but an implemented adaptive rule may use smaller bets. The formula applies only where $q_t<1$. It is not a universal envelope after clipping reaches one.

For $N=9378$, $u=0.1$, and the required wealth $L_1/\delta=640$, direct evaluation gives $U_{94}=165.632781$, $U_{118}=616.076398$, and $U_{119}=650.788159$. Hence 119 is the first sample size at which even the allowed zero-error envelope reaches 640. It is a necessary feasibility floor for a zero-error certificate under this cap, not sufficient for any particular aGRAPA implementation.

The earlier expression $\log M_n\leq n\,\mathrm{KL}(\hat\theta\|u)$ is not a general bound for this without-replacement process. For example, $N=10$, $u=0.1$, $n=8$, an all-zero sample, and cap bets give $\log U_8\approx0.854115$, exceeding $8\log(1/0.9)\approx0.842884$. Residual-population centering must be retained in the calculation.

\subsection{Extensions of the Certification Target}
\begin{remark}[Conditions for extending the target]
\label{rem:bounded}
The single-candidate proof extends to a fixed population of losses $\ell_i(\phi)\in[0,1]$: its conditional-mean identity and nonnegative-factor argument use boundedness, not binary labels. For a fixed finite family and a fixed within-track order, the first-error proof also controls false certification of these mean-loss constraints. Monotone losses are useful for the search interpretation but not required by that proof. However, an end-to-end guarantee additionally requires a fallback satisfying the new target, or separately certified with an allocated failure budget. Oracle fallback is zero-loss only for disagreement with that same oracle; arbitrary losses and ground-truth errors need not vanish there. Alternatively, the procedure may report no certified decision.

Class-conditional precision and recall are ratios with different target populations, not automatic instances of the overall-accuracy theorem. For a fixed subgroup of known size, independent of calibration sampling, conditioning on the number of sampled subgroup members gives a uniform without-replacement sample within it. The same test is then valid using that subgroup size and its random-order subsequence, conditional on a positive sample count. With zero sampled members, it must not certify a nontrivial bound from data. The zero-size target group also needs an explicit metric convention.

For recall, the positive-label population size is generally unknown without additional oracle information. For precision, the predicted-positive population changes with the candidate and may itself depend on oracle fallback labels. The current algorithm supplies neither all such population sizes nor simultaneous ratio guarantees. These extensions require an explicitly valid conditional-population or ratio procedure and appropriate error-budget allocation; they are not established here merely by substituting a loss or adding a Bonferroni split.
\end{remark}

\subsection{Cost Accounting with Cached Predictions}
\label{app:cost}
\noindent\textbf{Search overhead and prediction reuse.}
Equation~\eqref{eq:cost} prices execution of a selected plan without pre-existing prediction caches. It is distinct from the total cost of searching and executing on the same batch. If search has already evaluated Stage~1 on all records and Stage~2 on $\mathcal I_2^{\max}$, these predictions are reused. With $R(\phi)=\{i:D_i^{(1)}D_i^{(2)}=1\}$, same-batch modeled cost is
\begin{equation}
 C_{\rm batch}(\phi;S)=C_{\rm prep}+C_1+
 \sum_{i\in\mathcal I_2^{\max}}c_{2i}+\sum_{i\in S\cup R(\phi)}c_{Li},
 \label{eq:batch_cost}
\end{equation}
where $C_{\rm prep}$ accounts for retrieval, indexing, and search computation. Each invocation is charged once. Optimizing $C$ and optimizing $C_{\rm batch}$ can select different pairs. Our definition uses $C$; reported end-to-end costs must separately account for the actual cached execution. The certification argument also permits selecting by $C_{\rm batch}$ among certified candidates, but this is a different cost objective.

\subsection{Interpreting the Threshold-Selection Example}
\label{app:threshold-example}
The example in Section~\ref{sec:system} uses fixed scores and illustrative stage costs.
If this record's short-evidence prediction is wrong but its full-document prediction is correct, the second pair removes the error for three additional cost units. If both predictions are wrong, oracle fallback removes it. The reverse is also possible: the short-evidence prediction can be correct while the full-document prediction is wrong. More context does not guarantee a better answer.

Calibration tests aggregate disagreement under one shared threshold pair; it neither selects per-record thresholds nor requires every accepted answer to be correct. The cheapest pair is selected only among certified candidates. Cached predictions are reused, so the table describes plan cost rather than additional charges for cached execution.

\subsection{Notation}
\begin{table}[!h]

 \centering
 \caption{Primary notation. Certification threshold $\varepsilon = 1-\tau$.}
 \label{tab:notation}
 \footnotesize
 \setlength{\tabcolsep}{3.3pt}
 \begin{tabular}{@{}ll@{}}
 \toprule
 \headrow\textbf{Symbol} & \textbf{Meaning} \\
 \midrule
 $y_i$, $\hat{y}_i^{\Pi}$ & oracle label $\ML(q, d_i)$; prediction of plan $\Pi$ \\
 $\mathrm{cost}_S,\,\mathrm{cost}_L$ & token-aware invocation cost functions for $\MS$, $\ML$ \\
 $\tilde{z}_i^{(s)}$ & adaptive routing score at Stage $s$ (Eq.~\ref{eq:adaptive_score}) \\
 $\Lambda_1,\Lambda_2$ & Stage-1/2 threshold grids; $\Phi=\Lambda_1\times\Lambda_2$ \\
 $\phi=(\lambda_1,\lambda_2)$ & a candidate cascade configuration \\
 $\mathcal{I}_2^{\max}$ & maximal subset reaching Stage~2 (Eq.~\ref{eq:i2max}) \\
 $e_i(\phi)$, $E(\phi)$ & per-record error; population error rate $\frac1N\sum_i e_i(\phi)$ \\
 $r_3(\phi)$ & Stage-3 oracle-call ratio (Section~\ref{sec:guarantee}) \\
 $S$, $M_n^+$ & SRSWOR calibration sample ($|S|=n$); betting wealth \\
 $\mathcal{C}$, $\hat{\phi}$ & certified set; its cost-optimal member \\
 \bottomrule
 \end{tabular}
\end{table}

}
\end{document}